\documentclass[twocolumn]{aastex62}
\usepackage{xcolor}

\newcommand{\Kepler}{\textit{Kepler}}
\newcommand{\Gaia}{\textit{Gaia}}

\defcitealias{Mann2015}{M15}
\defcitealias{Dressing2013}{DC13}
\defcitealias{Dressing2015}{DC15}

\shorttitle{Mid-Type M Dwarf Planet Occurrence Rates}
\shortauthors{Hardegree-Ullman et al.}

\begin{document}

\title{\Kepler\ Planet Occurrence Rates for Mid-Type M Dwarfs as a Function of Spectral Type}

\correspondingauthor{Kevin K. Hardegree-Ullman}
\email{kevinkhu@caltech.edu}

\author[0000-0003-3702-0382]{Kevin K. Hardegree-Ullman}
\altaffiliation{Visiting astronomer, Kitt Peak National Observatory, National \\ Optical Astronomy Observatory, which is operated by the \\ Association of Universities for Research in Astronomy (AURA) \\ under a cooperative agreement with the National Science \\ Foundation. Visiting Astronomer at the Infrared Telescope \\ Facility, which is operated by the University of Hawaii under \\ contract NNH14CK55B with the National Aeronautics and \\ Space Administration.}
\affiliation{The University of Toledo, 2801 W. Bancroft St., Mailstop 111, Toledo, OH 43606 USA}
\affiliation{Caltech/IPAC-NExScI, MC 100-22, 1200 E. California Blvd., Pasadena, CA 91125 USA}

\author[0000-0001-7780-3352]{Michael C. Cushing}
\affiliation{The University of Toledo, 2801 W. Bancroft St., Mailstop 111, Toledo, OH 43606 USA}

\author[0000-0002-0638-8822]{Philip S. Muirhead}
\affiliation{Department of Astronomy \& The Institute for Astrophysical Research, Boston University, 725 Commonwealth Ave., Boston, MA 02215 USA}

\author[0000-0002-8035-4778]{Jessie L. Christiansen}
\affiliation{Caltech/IPAC-NExScI, MC 100-22, 1200 E. California Blvd., Pasadena, CA 91125 USA}



\begin{abstract}

Previous studies of planet occurrence rates largely relied on photometric stellar characterizations. In this paper, we present planet occurrence rates for mid-type M dwarfs using spectroscopy, parallaxes, and photometry to determine stellar characteristics. Our spectroscopic observations have allowed us to constrain spectral type, temperatures, and in some cases metallicities for 337 out of 561 probable mid-type M dwarfs in the primary \Kepler\ field. We use a random forest classifier to assign a spectral type to the remaining 224 stars. Combining our data with \Gaia\ parallaxes, we compute precise ($\sim$3\%) stellar radii and masses, which we use to update planet parameters and planet occurrence rates for \Kepler\ mid-type M dwarfs. Within the \Kepler\ field, there are seven M3\,V to M5\,V stars which host 13 confirmed planets between 0.5 and 2.5 Earth radii and at orbital periods between 0.5 and 10 days. For this population, we compute a planet occurrence rate of $1.19^{+0.70}_{-0.49}$ planets per star. For M3\,V, M4\,V, and M5\,V, we compute planet occurrence rates of $0.86^{+1.32}_{-0.68}$, $1.36^{+2.30}_{-1.02}$, and $3.07^{+5.49}_{-2.49}$ planets per star, respectively.
\end{abstract}



\section{Introduction}
\label{sec:intro}

The NASA \Kepler\ mission \citep{Borucki2010} revolutionized astrophysics and planetary science. \Kepler\ has enabled the discovery of 2,342 new exoplanets and an additional 2,421 planet candidates.\footnote{\url{http://exoplanetarchive.ipac.caltech.edu/docs/counts_detail.html}, as of April 2019} Four years of data from the original \Kepler\ mission have provided light curves for nearly 200,000 stars \citep{Mathur2017} from which we can estimate the statistical distribution of planet properties within our Galaxy.

\Kepler\ was designed to detect Earth-sized planets in the habitable zones of Sun-like (F, G, and K) stars. M dwarfs ($2,300\,\mathrm{K} \lesssim T_{\mathrm{eff}} \lesssim 3,900\,\mathrm{K}$, $0.1\,R_{\odot} \lesssim R_{\star} \lesssim 0.6\,R_{\odot}$, $0.07\,M_{\odot} \lesssim M_{\star} \lesssim 0.6\,M_{\odot}$) comprise about $70\%$ of the nearby stellar population, by number \citep{Henry2006,Bochanski2010}, though they only constitute about 2.5\% of the targets \Kepler\ observed \citep[][hereafter \citetalias{Dressing2015}]{Dressing2015}. The smaller size of these stars makes it easier to detect the presence of smaller transiting planets. For example, the transit depth for an Earth-radius planet transiting a solar metallicity M0\,V star would be 3 times deeper than the same planet transiting a G2\,V Sun-like star, and nearly 70 times deeper for an M7\,V star. Fortuitously, the low luminosity of M dwarfs means the habitable zone is closer to the star, thus increasing the chance of detecting a transiting planet in the habitable zone over a finite observing period \citep{Nutzman2008}. For example, an Earth-sized planet in the habitable zone of an M4\,V star orbits once every two weeks as opposed to once a year for a Sun-like G star.

Planet occurrence rates increase toward later spectral types within the \Kepler\ field. Using the first three quarters of \Kepler\ data, \citet{Howard2012} measured the planet occurrence rate for M0 to F2 dwarfs for planets with radii between 2 and $4\,R_{\oplus}$ and found that these small planets are seven times more abundant around cool stars (3,600 to $4,100\,\mathrm{K}$) than hot stars (6,600 to $7,100\,\mathrm{K}$). More recent work by \citet{Mulders2015} found the occurrence rate of planets with radii between 1 and $4\,R_{\oplus}$ around M dwarfs to be two times higher than for G stars, and three times higher than for F stars. \citet[][hereafter \citetalias{Dressing2013}]{Dressing2013} focused specifically on stars with temperatures below $4,000\,\mathrm{K}$ in Q1 to Q6 of \Kepler\ data and found the occurrence rate of 0.5 to $4\,R_{\oplus}$ planets with orbital periods shorter than 50 days to be $0.90^{+0.04}_{-0.03}$ planets per star. Separating these stars into warmer ($T_{\mathrm{eff}} > 3,400\,\mathrm{K}$) and cooler ($T_{\mathrm{eff}} < 3,400\,\mathrm{K}$) groups, they find the occurrence rate of Earth-sized planets (0.5 to $1.4\,R_{\oplus}$) to be consistent at around 0.5 planets per star, but the rate of larger planets (1.4 to $4.0\,R_{\oplus}$) is three times higher for the warmer stars, $0.61^{+0.08}_{-0.06}$ compared to $0.19^{+0.07}_{-0.05}$. Using the full \Kepler\ data set, \citetalias{Dressing2015} updated their M dwarf planet occurrence rate for 1 to $4\,R_{\oplus}$ planets with orbital periods shorter than 200 days to be $2.5\pm0.2$ planets per star, but they do not make a distinction between early and mid-type M dwarfs. \citet{Gaidos2016b} computed an overall \Kepler\ M dwarf planet occurrence rate of $2.2\pm0.3$ for orbital periods shorter than 180 days, consistent with \citetalias{Dressing2015}. Focusing specifically on mid-type M dwarfs, \cite{Muirhead2015} calculated a compact multiple occurrence rate of $21^{+7}_{-5}\%$, assuming a radius of $0.2\,R_{\odot}$ for all mid-type M dwarfs.

Computing planet occurrence rates requires measurements of stellar radii ($R_{\star}$) for all stars in the sample population. In the absence of direct measurements, $R_{\star}$ for the M dwarfs in the \Kepler\ Input Catalog (KIC) was derived from optical and infrared photometry \citep{Batalha2010,Brown2011}. \citet{Brown2011} noted that stellar parameters in the KIC are unreliable at $T_{\mathrm{eff}} < 3,750\,\mathrm{K}$ because the models they use are calibrated to work best for Sun-like stars. \citetalias{Dressing2013}, \citet{Gaidos2013}, and \citet{Huber2014} addressed this issue by using stellar models more suited to later type stars \citep[e.g., Dartmouth Stellar Evolution Database;][]{Dotter2008} to reclassify the set of cooler stars in the KIC. These updated measurements of $R_{\star}$, based on photometry, still have a wide range of uncertainties with an average around 30\% \citep{Mathur2017}. An uncertainty this large can mean the difference between several spectral sub-types, and for a transiting planet this propagates to a 30 to 40\% planet radius uncertainty. 

Combining long-baseline optical interferometry with trigonometric parallax measurements yields direct measurements of stellar radii with uncertainties between 1\% and 5\%; however, this method is currently limited to bright ($V \lesssim 11$), nearby stars \citep[e.g.,][]{Segransan2003,Boyajian2012,vonBraun2014}. These precise radius measurements can, however, be used to calibrate empirical relationships between radius and other measurable parameters \citep[e.g.,][]{Boyajian2012,Mann2015}. For example, \citet[][hereafter \citetalias{Mann2015}]{Mann2015} derive $M_{K_s}$ vs.\ $R_{\star}$ and $T_{\mathrm{eff}}$ vs.\ $R_{\star}$ relationships for K7 through M7 dwarfs, which constrain stellar radii to $\sim$3\% and $\sim$15\%, respectively. \citet{Gaidos2016b} used these relationships to update the stellar properties of over 4,000 \Kepler\ M dwarfs based on temperatures, metallicities, and distances derived from photometry and proper motions. These measurements constrain stellar radii to $\sim$15\%.

With moderate resolution spectra, we can precisely determine spectral type and measure stellar properties such as temperature and metallicity. Having accurate spectral types for a specific population (e.g., mid-type M dwarfs) allows us to assess trends in planet occurrence rates for that population as a function of spectral type, like the assessment of F, G, K, and early M dwarf planet occurrence rates by \citet{Howard2012}. In large stellar catalogs, such as the KIC, it is difficult to obtain a spectrum of each star, so photometric selection criteria are typically used to identify specific stellar populations \citep[e.g.,][]{Brown2011,Dressing2013,Gaidos2013,Huber2014,Gaidos2016b}. Efforts to spectroscopically classify nearly 5,000 photometrically-identified late-type K and M dwarfs in the \Kepler\ field by \citet{Mann2012}, \citet{Muirhead2012}, \citet{Mann2013b}, and \citet{Martin2013} have yielded spectra for only $2\%$ of the sample \citepalias{Dressing2015}. While selection criteria aim to minimize stellar outliers, they are not impervious to contamination. Spectra allow us to identify any interlopers (such as giant stars) that selection criteria might not cull.

In this paper we present spectroscopic observations of 333 M dwarfs and 4 M giants in the \Kepler\ field, from which we derive spectral types, effective temperatures, and metallicities. Thanks to the European Space Agency's \Gaia\ mission \citep{Gaia2016}, nearly all the stars in the \Kepler\ field now have trigonometric parallax measurements \citep{Berger2018}. Over $90\%$ of our targets have \Gaia\ parallax measurements, which we use to compute $M_{K_s}$. We apply the empirical relationships of \citetalias{Mann2015} and \citet{Mann2019} to measure radii and masses for our stars. With these updated stellar properties, we refine the properties for the 13 confirmed planets around mid-type M dwarfs. We then compute the total planet occurrence rate for mid-type M dwarfs, individual occurrence rates for M3\,V, M4\,V, and M5\,V stars, and compact multiple occurrence rates.

\begin{figure*}
\gridline{\fig{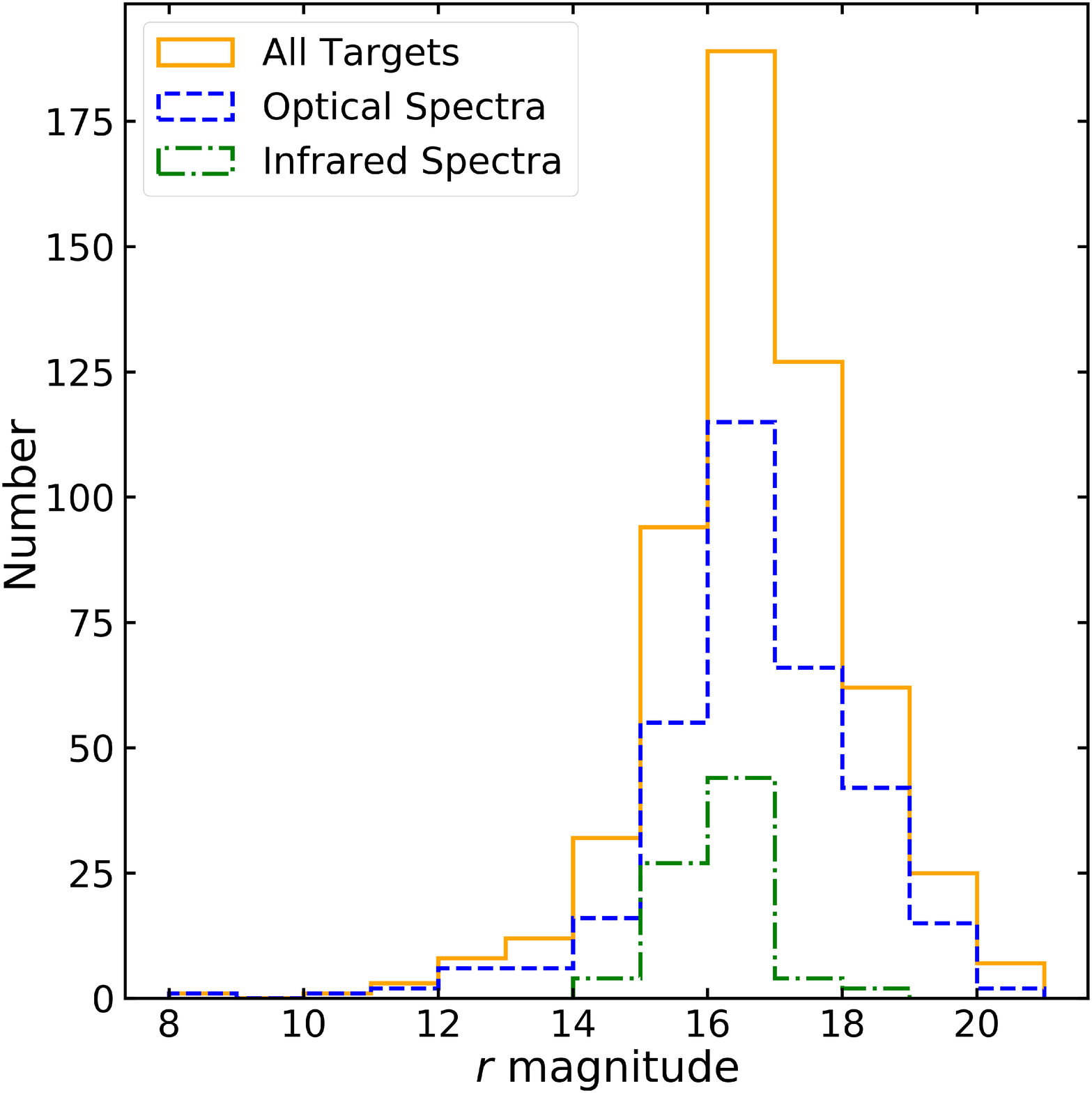}{0.47\textwidth}{\ \ \ \ \ \ \ \ \ \ (a)}
          \fig{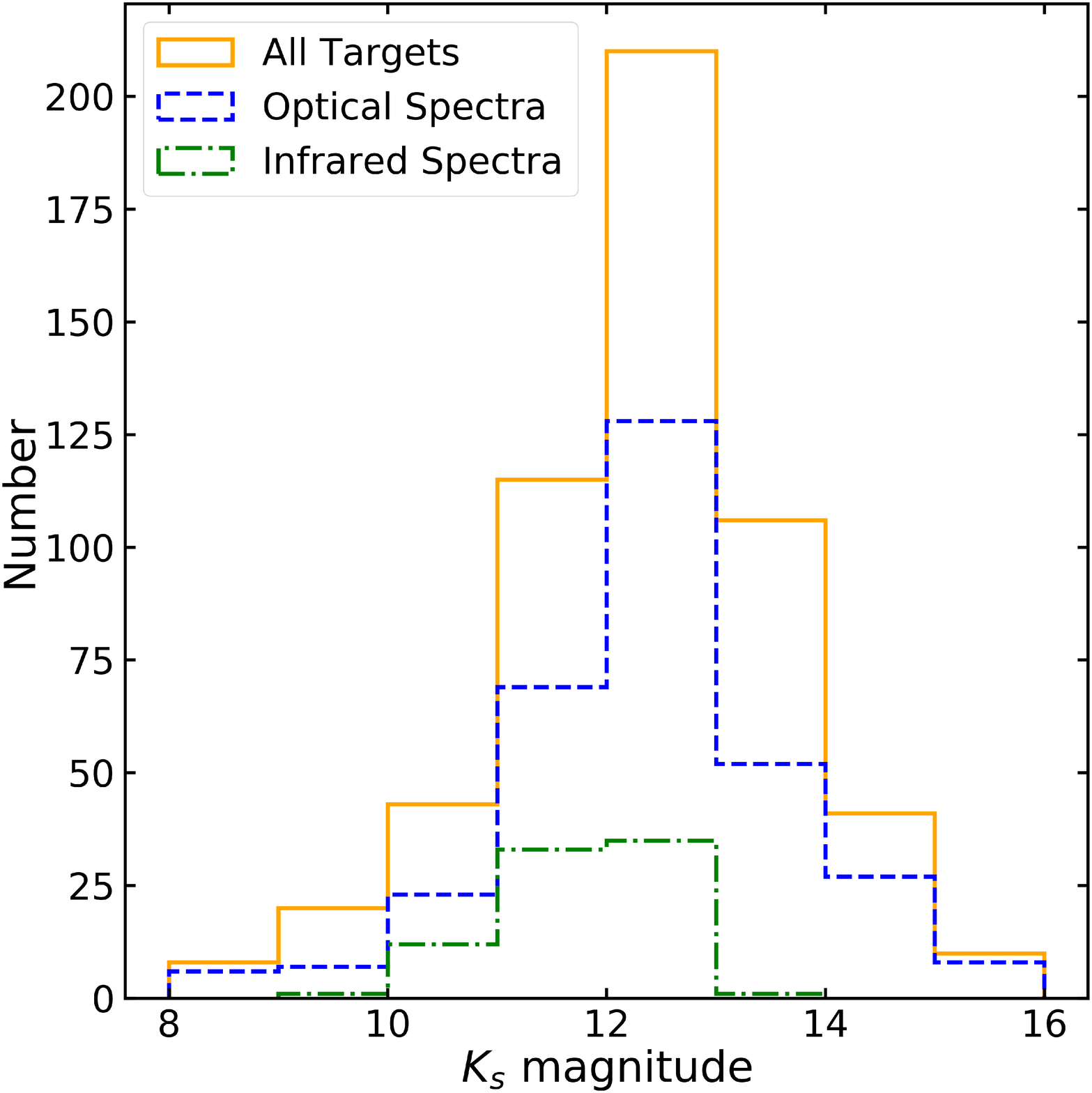}{0.47\textwidth}{\ \ \ \ \ \ \ \ \ \ (b)}}
\caption{Magnitude distributions in (a) $r$-band and (b) $K_s$-band of \Kepler\ targets for this survey. The orange solid line histograms represent all targets, the blue dashed line histograms are the targets for which we have an optical spectrum, and the red dashed dotted line histograms show targets for which we have an infrared spectrum. The optical spectra distribution closely follows the distribution of all targets in the sample. Targets were selected to optimize observation efficiency with WIYN by maximizing the number of targets observed per field configuration. Several fainter targets were also observed with the Discovery Channel Telescope. This strategy allowed us to effectively observe a random sample representing the total target set. Almost all stars with infrared spectra have $K_s < 13$, since fainter targets necessitate much longer integration times to obtain sufficient S/N to measure stellar properties.}\label{fig:spechist}
\end{figure*}

In Section~\ref{sec:observations} we describe our target selection, spectroscopic observations, and \Gaia\ data. We explain our methods for determining stellar properties in Section~\ref{sec:props}, and refine our sample to reject giant stars and close binaries in Section~\ref{sec:gb}. In Section~\ref{sec:planets} we identify the planets around mid-type M dwarfs in the \Kepler\ field, which we use for our planet occurrence rate calculations in Section~\ref{sec:occur}. We conclude with a discussion in Section~\ref{sec:discussion}.

\section{Target Selection, Observations, and Data Reduction}
\label{sec:observations}

Using revised KIC temperature estimates, \citetalias{Dressing2013} identified 202 stars with $T_{\rm{eff}} \lesssim 3,300\,\mathrm{K}$. \citet{Muirhead2015} identified 509 probable mid-type M dwarfs from KIC photometry using the following color selection criteria: $r-J > 3.2$ to identify red objects, and $J-K < 0.0555 \times (r-J) + 0.7622$ to remove giant stars. The union between these samples produces 561 probable mid-type M dwarfs which we targeted in this study. In total, we observed 337 of the 561 targets (60\%), obtaining optical spectra of 327 stars and near-infrared spectra of 82 stars. We have both optical and infrared spectra for 72 targets. Magnitude distributions of our targets in $r$-band and $K_s$-band are shown in Figure~\ref{fig:spechist}, including the distributions of targets we observed.

\subsection{WIYN}
\label{sec:wiyn}

In observing semester 2015B, NASA and NSF implemented Stage 1 of the Exoplanet Observational Research (NN-EXPLORE) program, enabling community access to about 100 nights per year on the 3.5-meter WIYN\footnote{The WIYN Observatory is a joint facility of the University of Wisconsin-Madison, Indiana University, the National Optical Astronomy Observatory and the University of Missouri.} telescope on Kitt Peak at least through commissioning of the extreme precision Doppler spectrometer NEID \citep{Schwab2016}. We have made use of the fiber-fed multi-object spectrograph Hydra \citep{Barden1994} on WIYN for this project. The \Kepler\ field spans a 12 degree diameter, making the one degree diameter field-of-view of Hydra advantageous. We collected spectra over five observing semesters beginning in September 2015 (NOAO Program IDs 2015B-0280, 2016A-0328, 2016B-0111, 2017A-0185, 2017B-0095; PI: K. Hardegree-Ullman). In total, we observed 287 targets with WIYN.

Each Hydra field configuration contained an average of five mid-type M dwarf targets with $V < 18.5$\footnote{KIC photometry was converted to $V$-band magnitudes using the transformation $V=g-0.5784\times(g-r)-0.0038$, \url{http://www.sdss3.org/dr8/algorithms/sdssUBVRITransform.php\#Lupton2005}}. We used the bench spectrograph camera and the red optimized fiber cable containing 90 fibers. The 316~lines~mm$^{-1}$ grating and the G5 filter were used to obtain spectra spanning 5,000 to 10,000$\,$\AA\ with an average resolving power $R=\lambda/\Delta\lambda\approx1,420$. Most fields were observed in three consecutive 1,200 second exposures to mitigate the effect of cosmic rays and to yield a signal-to-noise (S/N) of at least 30 per resolution element for targets with $V < 18.5$. At the beginning and end of each night we obtained bias, dome flat field, and copper-argon wavelength calibration frames. After each field observation, we obtained a spectrum of a nearby bright G0 star at a similar airmass to be used for telluric correction. At least once each night we observed a spectrophotometric standard star for relative flux calibration. 

The data were reduced using IRAF\footnote{IRAF is distributed by the National Optical Astronomy Observatories, which are operated by the Association of Universities for Research in Astronomy, Inc., under cooperative agreement with the National Science Foundation.} and custom IDL routines. Each image was first corrected for readout bias using the overscan strip, then trimmed to remove this region. An average bias was then created and subtracted from the rest of the images. The \texttt{dohydra} IRAF routine was used for aperture extraction, flat fielding, dispersion correction, and sky subtraction. Sky spectra were created from 5 to 10 sky fibers placed in random positions distributed throughout each field. Sky subtraction removed OH atmospheric emission, which is problematic at wavelengths redder than 6,000$\,$\AA. Individual exposures were then combined, and a relative flux calibration was performed on all targets using a spectrophotometric standard star. We developed a custom IDL script to interpolate across a G0 star spectrum where there is known atmospheric absorption, and divide these regions out of our M dwarf spectra to remove terrestrial atmospheric absorption features. We shift each spectrum in wavelength to the source star's rest frame by cross-correlation with the closest matching M dwarf template from \citet{Bochanski2007}. We find the closest matching M dwarf template from a by-eye comparison of each spectrum to each template, normalized to 8,350$\,$\AA.

\vspace{10pt}
\subsection{Discovery Channel Telescope}
\label{sec:dct}

About 90 of our targets have $V$-band magnitudes fainter than 18.5, which would require integration times longer than one hour using WIYN to achieve a S/N of 30. We therefore used the DeVeny spectrograph on the 4.3-meter Discovery Channel Telescope (DCT) to observe some of these fainter targets over four observing runs, collecting spectra of 49 stars. We observed 17 of these stars with WIYN because they were within one of the observed WIYN fields. These targets are used to check for consistency between the two telescopes, though the DCT data are higher resolution.

The DeVeny spectrograph is a single-slit instrument with a deep depletion e2v CCD. The 400~grooves~mm$^{-1}$ grating and OG570 filter were used to obtain spectra spanning 6,300 to 9,700$\,$\AA\ with $R\approx2,850$. Total integration times varied in order to achieve a S/N similar to WIYN targets, and at least three consecutive exposures were taken of each target to mitigate cosmic rays in the data reduction. Calibrations were the same as for WIYN, except for the neon-argon lamps used for wavelength calibration. Data reduction was again performed in IRAF using the same steps as for WIYN data, but with single slit routines instead of \texttt{dohydra}. 

\subsection{IRTF}
\label{sec:irtf}

We used the SpeX spectrograph \citep{Rayner2003} on the NASA Infrared Telescope Facility (IRTF) on Maunakea to observe 82 targets over 12 partial nights (Program IDs 2016A-981, 2017A-106, 2017B-021; PI: K. Hardegree-Ullman). We used SpeX in SXD mode with a $0\farcs3\times15\arcsec$ slit to obtain 0.8 to $2.4\,\mu$m spectra at $R\approx2,000$. We observed using the ABBA nod pattern with at least three exposures for each target. Exposure times were shorter than 120 seconds to minimize effects from atmospheric OH line variability. For most targets we achieved a S/N of at least 50 per resolution element. An A0\,V star was observed within 0.1 airmasses of each target to be used for telluric correction and flux calibration. Flat field and thorium-argon lamp calibrations were taken after each A0\,V star observation.

Spectra were reduced using \texttt{Spextool} \citep[v.\ 4.2;][]{Cushing2004}, which performs flat fielding, wavelength calibration, sky subtraction, and spectrum extraction. We used the \texttt{xtellcor} IDL package \citep{Vacca2003} for telluric correction. As with each optical spectrum, we shift each infrared spectrum in wavelength to the source star's rest frame by cross-correlating with a spectrum of corresponding spectral type from the IRTF spectral library \citep{Cushing2005,Rayner2009}.

\vspace{20pt}
\subsection{LAMOST}
\label{sec:lamost}

The Large Sky Area Multi-Object Fiber Spectroscopic Telescope \citep[LAMOST;][]{Cui2012} is a 4-meter class telescope designed to survey stars and galaxies in the northern hemisphere. Its 4,000 fiber multi-object spectrograph has a wavelength range of 3,690 to 9,100$\,$\AA\ at $R\approx1,800$. As of data release 4, over 7.6 million\footnote{\href{http://dr4.lamost.org/}{http://dr4.lamost.org/}} spectra have been gathered in total. There are LAMOST spectra for 17 of our \Kepler\ targets, including 8 targets we did not observe with WIYN or the DCT. This brings our total optical spectra count to 327 targets.

\subsection{Gaia}
\label{sec:gaia}

The \Gaia\ mission has provided parallax measurements for over 1.3 billion sources in second data release \citep{Gaia2018}. \citet{Bailer-Jones2015} and \citet{Bailer-Jones2018} note that simply inverting the parallax does not always produce reliable distances. Instead, distances must be inferred via probabilistic analysis, which \citet{Bailer-Jones2018} performed on all targets in the second \Gaia\ data release, providing distance measurements and uncertainties. We use the \Gaia/\Kepler\ cross-match database\footnote{\href{http://gaia-kepler.fun/}{http://gaia-kepler.fun/}} to obtain the \citet{Bailer-Jones2018} \Gaia\ distances to our \Kepler\ stars. In total, 532 of our targets cross-match with a \Gaia\ target within 4\arcsec, the \Kepler\ pixel size. We also check the difference between the \Gaia\ $G$-band \citep[$\gtrsim 20\%$ transmission between 4,000 and 9,000$\,$\AA;][]{Evans2018} and \Kepler\ $Kp$-band ($\gtrsim 20\%$ transmission between 4,300 and 8,900$\,$\AA\footnote{\url{https://keplergo.arc.nasa.gov/CalibrationResponse.shtml}}) magnitudes. For any target with an absolute $G-Kp$ magnitude difference greater than 2, we discard the parallax measurement in our analysis, which applies to KIC~02164791, KIC~03330684, KIC~07729309, and KIC~10665619.

\section{Stellar Properties}
\label{sec:props}

We seek to determine planet occurrence rates for mid-type M dwarfs using revised stellar radius measurements derived from spectra, photometry, and parallaxes. About 95\% of our targets have parallax measurements, which allows us to compute absolute magnitudes in the $r$, $J$, and $K_s$-bands for use in computing stellar properties (Section~\ref{sec:absmag}). For the 337 targets with spectra, we do a by-eye comparison to spectral templates to determine spectral types. There are 224 targets for which we do not have spectra, so we use a machine learning technique to identify spectral type from photometry (Section~\ref{sec:spt}). Using a spectral type vs.\ temperature relationship, we derive stellar temperatures (Section~\ref{sec:teff}). For targets with infrared spectra, we determine metallicity (Section~\ref{sec:feh}). We apply the $M_{K_s}$ vs.\ radius and $M_{K_s}$ vs.\ mass relationships of \citetalias{Mann2015} and \citet{Mann2019} to derive stellar radii and masses for targets with parallax measurements. For the remaining targets, we derive radii from a temperature vs.\ radius relationship (Section~\ref{sec:rstar}) and masses from a radius vs.\ mass relationship (Section~\ref{sec:mstar}). With our updated spectral types, we isolate mid-type M dwarfs, and compute planet occurrence rates for spectral types M3\,V to M5\,V using our new stellar properties (Section~\ref{sec:occur}). We present all derived stellar parameters in Tables~\ref{tab:spec} and~\ref{tab:phot}.

\subsection{Absolute Magnitudes}
\label{sec:absmag}

For targets with \Gaia\ parallax measurements, we compute absolute magnitudes $M$ for Sloan Digital Sky Survey \citep[SDSS;][]{York2000} $r$-band and Two Micron All-Sky Survey \citep[2MASS;][]{Skrutskie2006} $J$ and $K_S$-bands using
\begin{equation}
    M=m-5[\log_{10}(d)-1]-A,
\end{equation}
where $m$ is the apparent magnitude for a photometric band, $d$ is the distance in parsecs, and $A$ is the extinction for that photometric band. We compute reddening vectors using the 3D dust maps of \citet{Green2018} with the Python package \texttt{dustmaps}. We find extinction for each band by multiplying the reddening vectors by the extinction coefficients $R$ from \citet{Green2018}, which are 2.483, 0.650, and 0.161 for $r$, $J$, and $K_s$ bands, respectively. We calculate absolute magnitudes and their associated uncertainties for each band using the following Monte Carlo (MC) analysis. For each measured parameter we generate $10^4$ random samples from a Gaussian distribution and combine these distributions for each calculation. The uncertainties for each quantity are symmetric or nearly symmetric within a few percent, so we use the mean of any asymmetric uncertainty to draw our Gaussian distribution. The absolute magnitude is then the median value of the posterior distribution, and we adopt the 16$^{\mathrm{th}}$ and 84$^{\mathrm{th}}$ percentiles as the uncertainties.

\subsection{Spectral Type}
\label{sec:spt}

We determine optical spectral types by comparing our spectra to the SDSS M dwarf spectral templates of \citet{Bochanski2007}, following the methods of \citet{Kirkpatrick1991}. The template spectra span a wavelength range of 3,800 to 9,200$\,$\AA\ with $R\approx1,800$. Target spectra are normalized to the same wavelength as the template spectra (8,350$\,$\AA) and we do a by-eye comparison to find the closest matching spectrum to the nearest half spectral type. Ten of our targets were only observed with IRTF. For these targets, we determine spectral type in the same manner as our optical spectra, except we use M dwarfs in the IRTF Spectral Library for comparison \citep{Cushing2005, Rayner2009}.

\begin{figure}[ht!]
    \centering
    \includegraphics[width=0.47\textwidth]{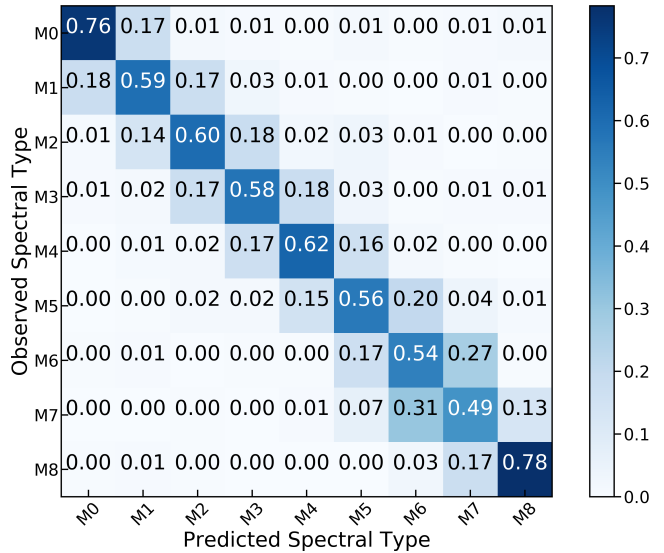}
    \caption{Comparison of the fraction of predicted classifications in each spectral type bin versus observed spectral type from a random forest classification using photometry from the \citet{West2011} M dwarf sample. Over 90\% of the predicted classifications are within one spectral type of the observed spectral type.}\label{fig:trainrf}  
		\vspace{0em}
\end{figure}

It is always best to determine spectral type from a spectrum rather than photometry, however, this is not always economical with a large target sample and limited resources. For the 224 stars in our sample that lack a spectrum, we use the \texttt{RandomForestClassifier} routine from the \texttt{scikit-learn} \citep{Pedregosa2011} Python package to classify these targets with $r$, $J$, and $K_s$-band photometry. For our training sample, we use the extinction corrected photometry and spectral types from \citet{West2011}, who visually inspected over 70,000 M dwarf spectra from SDSS. We truncate this data set to include spectral types M0 to M8, and use the color cut $J-K_s < 0.0555 \times (r-J) + 0.7622$ to reflect the targets in our sample. From these 48,978 targets, we randomly select 1,500 from each spectral type to minimize sample bias, reducing our input sample to 13,500 targets. Using the $r-J$ and $J-K_s$ colors along with spectral type, we train the random forest classifier on a random subset of 75\% of the total input sample. The remaining 25\% of the input sample is used to determine how well the classifier performs. Figure~\ref{fig:trainrf} compares the observed spectral type from spectra to the predicted spectral type from the classifier. Over 90\% of the predicted classifications are within one spectral type of the observed spectral type, so we adopt an uncertainty of $\pm 1$ spectral type for our photometric classifications.

\begin{figure*}
\gridline{\fig{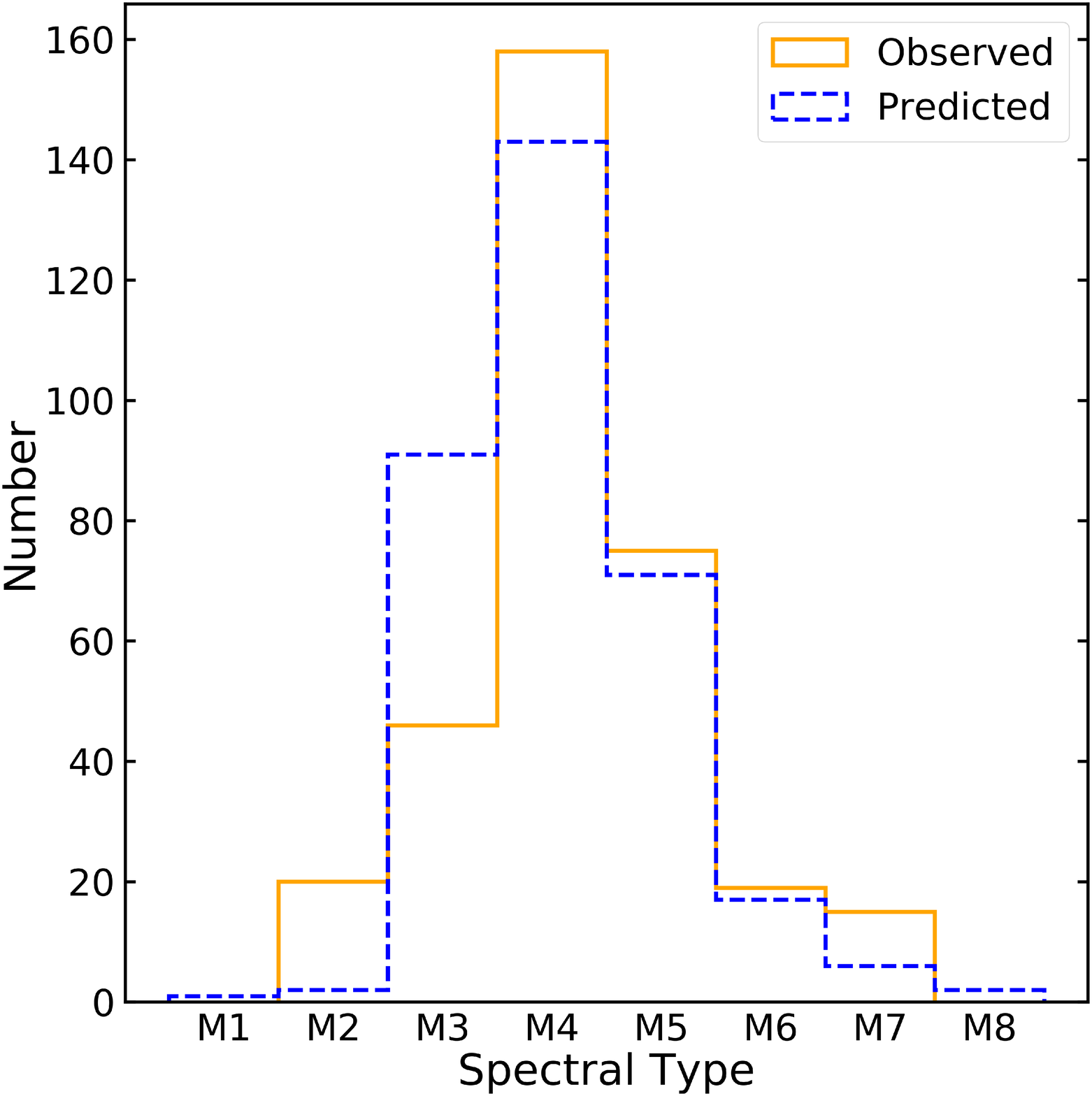}{0.47\textwidth}{\ \ \ \ \ \ \ \ \ \ (a)}
          \fig{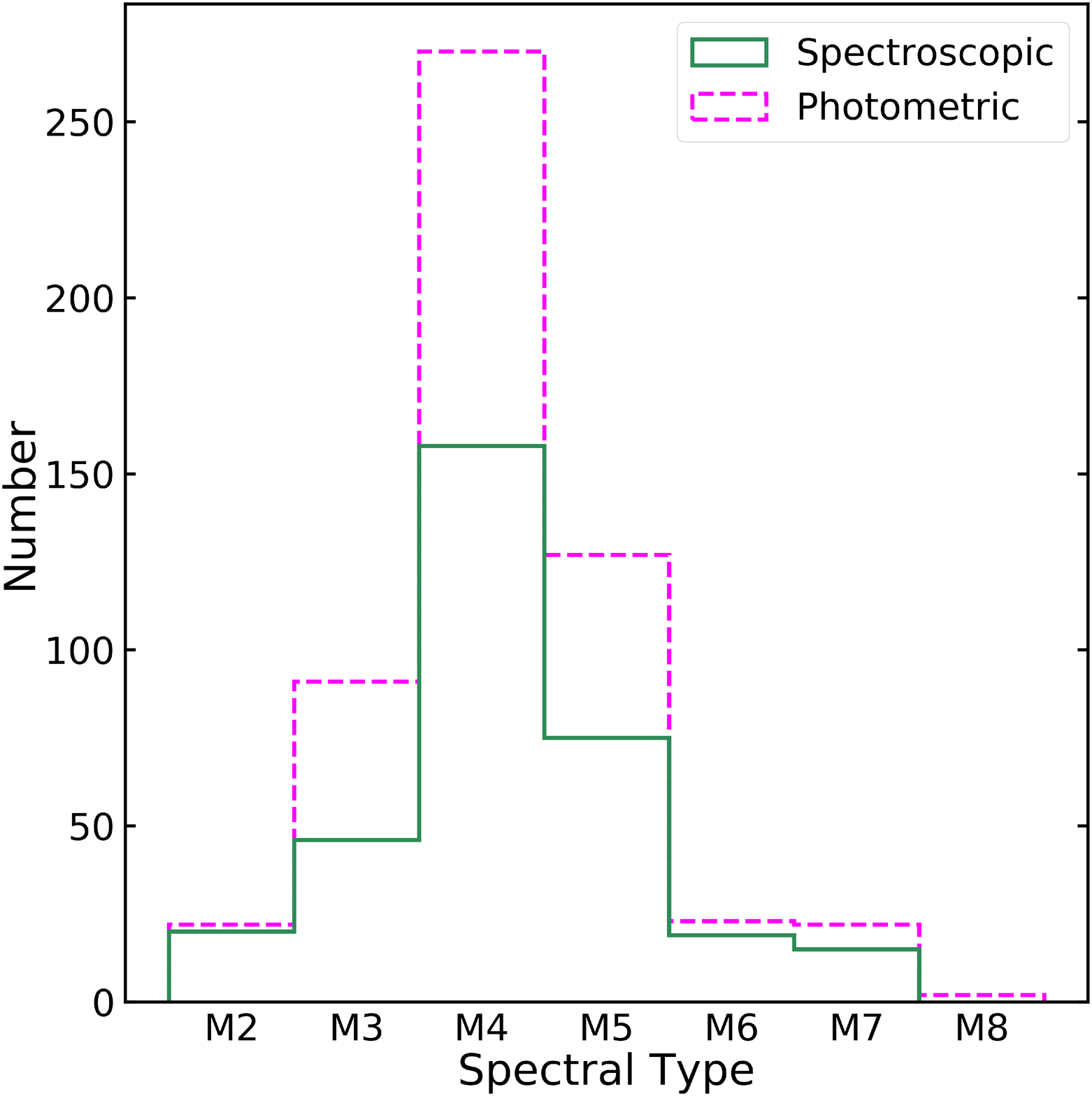}{0.47\textwidth}{\ \ \ \ \ \ \ \ \ \ (b)}}
\caption{(a) Comparison of observed (solid orange) versus predicted (dashed blue) spectral type distributions using the random forest classifier on the \Kepler\ targets we have spectroscopically classified. (b) The spectral type distribution for our entire \Kepler\ sample using results from spectra for observed targets (solid green), and the classifier (dashed magenta) on stars without spectra. The photometric histogram is stacked on top of the spectroscopic histogram. For these histograms, stars with half spectral types are rounded up to the nearest integer type.}\label{fig:classrf}
\end{figure*}

We run the trained classifier on the photometry for both our spectroscopically observed and unobserved targets. In the case of the 530 targets for which we have \Gaia\ parallax measurements, we use extinction corrected absolute magnitudes to compute $r-J$ and $J-K_s$ colors. The average difference between the colors computed from absolute magnitudes versus apparent magnitudes is about 0.03 for each color, which is the small effect of the reddening correction. For the remaining 31 targets without parallax measurements, we use apparent magnitudes to compute colors. The classifier predicts 86\% of the spectral types of our spectroscopically observed sample to within one spectral type, however, it over-predicts M3\,V by a factor of 2, and significantly under-predicts M2\,V and M7\,V, as shown in Figure~\ref{fig:classrf}~a. For the photometric sample, the distribution of spectral types roughly follows the same distribution as the spectroscopic sample. We also show the spectral type distribution for the entire sample in Figure~\ref{fig:classrf}~b. Since we effectively observed a random subset of mid-type M dwarfs in the \Kepler\ field, we expect the photometrically classified targets to follow a similar spectral type distribution to the spectroscopically classified targets, and thus we are confident in the results from the photometric classification.

\subsection{Effective Temperature}
\label{sec:teff}

Direct measurement of effective temperatures requires both bolometric flux and interferometric angular diameter measurements. Due to the distance and faintness of these stars, interferometric observations are not currently possible. Using the spectral types and effective temperatures of the 183 stars in Table~5 of \citetalias{Mann2015}, we determine a linear spectral type vs.\ effective temperature relationship:
\begin{equation}
    T_{\mathrm{eff}}=-168.37 \times SpT+3914.65,
\end{equation}
where $SpT$ is the spectral type between $-1$ for K7\,V and 7 for M7\,V. We adopt this as a uniform temperature scale across our spectroscopic and photometric sample. The residual 1-$\sigma$ scatter of our fit to the \citetalias{Mann2015} spectral types and temperatures is $64\,\mathrm{K}$, which we add in quadrature to the typical $60\,\mathrm{K}$ spectroscopic measurement uncertainties. This yields $88\,\mathrm{K}$, which we adopt as our temperature uncertainty.

\citet{Pecaut2013} determine effective temperatures for each sub-type of main sequence stars by taking the median values of published effective temperatures in the literature. As a consistency check to our \citetalias{Mann2015} fit, we take the most recent list of all reported values used to compute the median effective temperature for each sub-type\footnote{\href{http://www.pas.rochester.edu/~emamajek/spt/}{http://www.pas.rochester.edu/$\sim$emamajek/spt/}} between K7\,V and M9\,V (280 stars) and fit a linear relationship. This results in $T_{\mathrm{eff}}=-166.96\times SpT + 3868.39$, with a $107\,\mathrm{K}$ 1-$\sigma$ scatter. This relationship gives temperatures on average $40\,\mathrm{K}$ cooler than our temperatures. We also compare our temperatures to the M dwarf temperature scale of \citet{Rajpurohit2013}, which are on average $76\,\mathrm{K}$ cooler than our temperatures. All three methods yield temperatures which are consistent within the measurement uncertainties.

\subsection{Metallicity}
\label{sec:feh}

We computed metallicities (specifically, iron abundance, [Fe/H]) for 82 targets we observed with SpeX on IRTF. \citet[][]{Mann2013a, Mann2014} determined metallicity of M dwarfs in wide binary systems with an F, G, or K primary star, and computed empirical relationships between metallicity and equivalent widths of Na I and Ca I using infrared spectra from SpeX. Our metallicity measurements using these relationships are listed in Table~\ref{tab:spec}. We did not obtain infrared spectra of any of the mid-type M dwarf planet hosts (Table~\ref{tab:planets}), however, we adopt the metallicity measurements of Kepler-42, Kepler-445, and Kepler-1650 from \citet{Mann2017}, Kepler-446 and Kepler-1582 from \citet{Muirhead2014}, and Kepler-1649 from \citet{Angelo2017}. All of these metallicity measurements were made using the same relationships from \citet{Mann2013a}. The only planet host for which we do not have a metallicity measurement is Kepler-1646.

\begin{figure*}[htp!]
\gridline{\fig{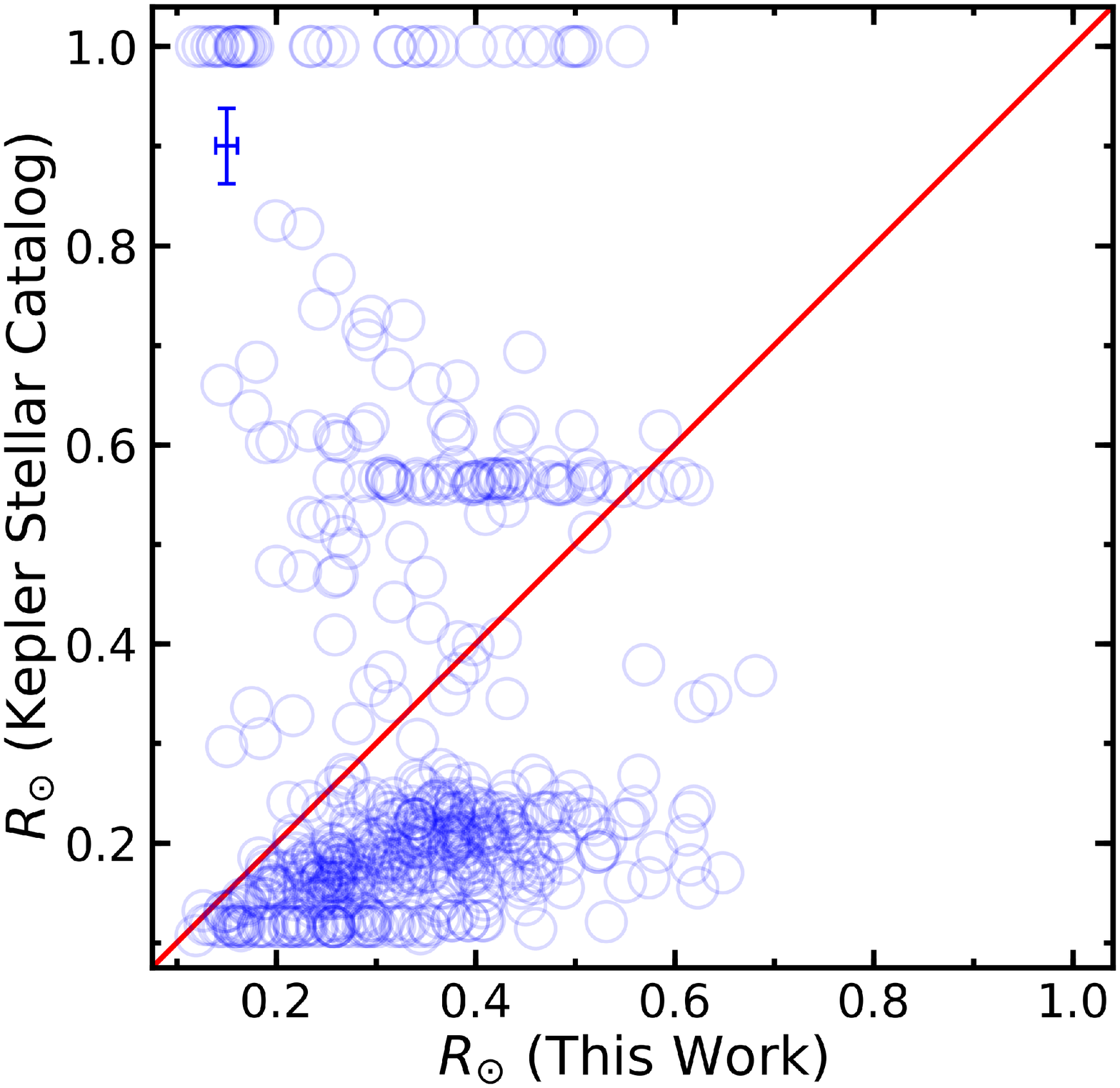}{0.3\textwidth}{\ \ \ \ \ \ \ \ (a)}
          \fig{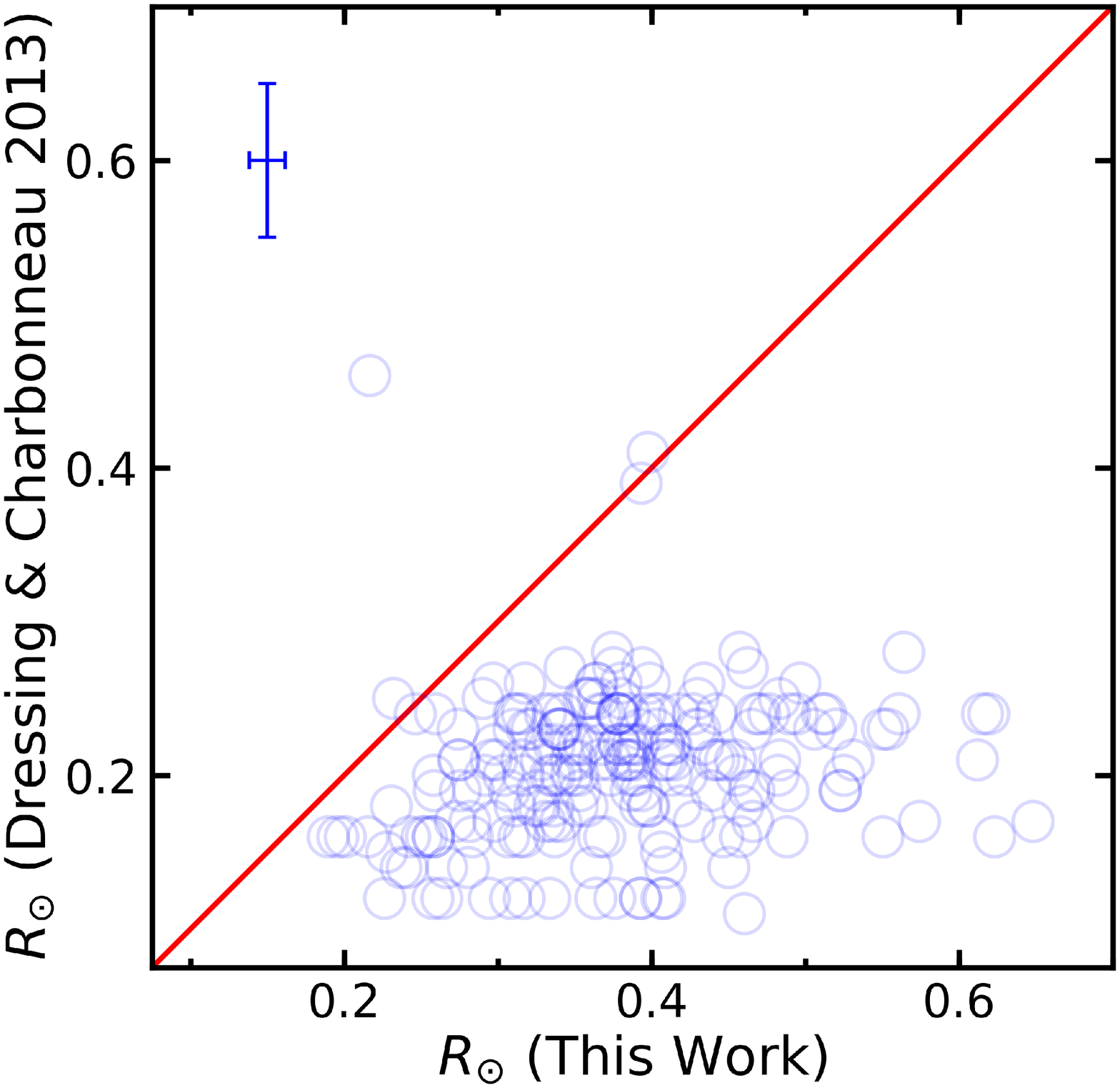}{0.3\textwidth}{\ \ \ \ \ \ \ \ (b)}
          \fig{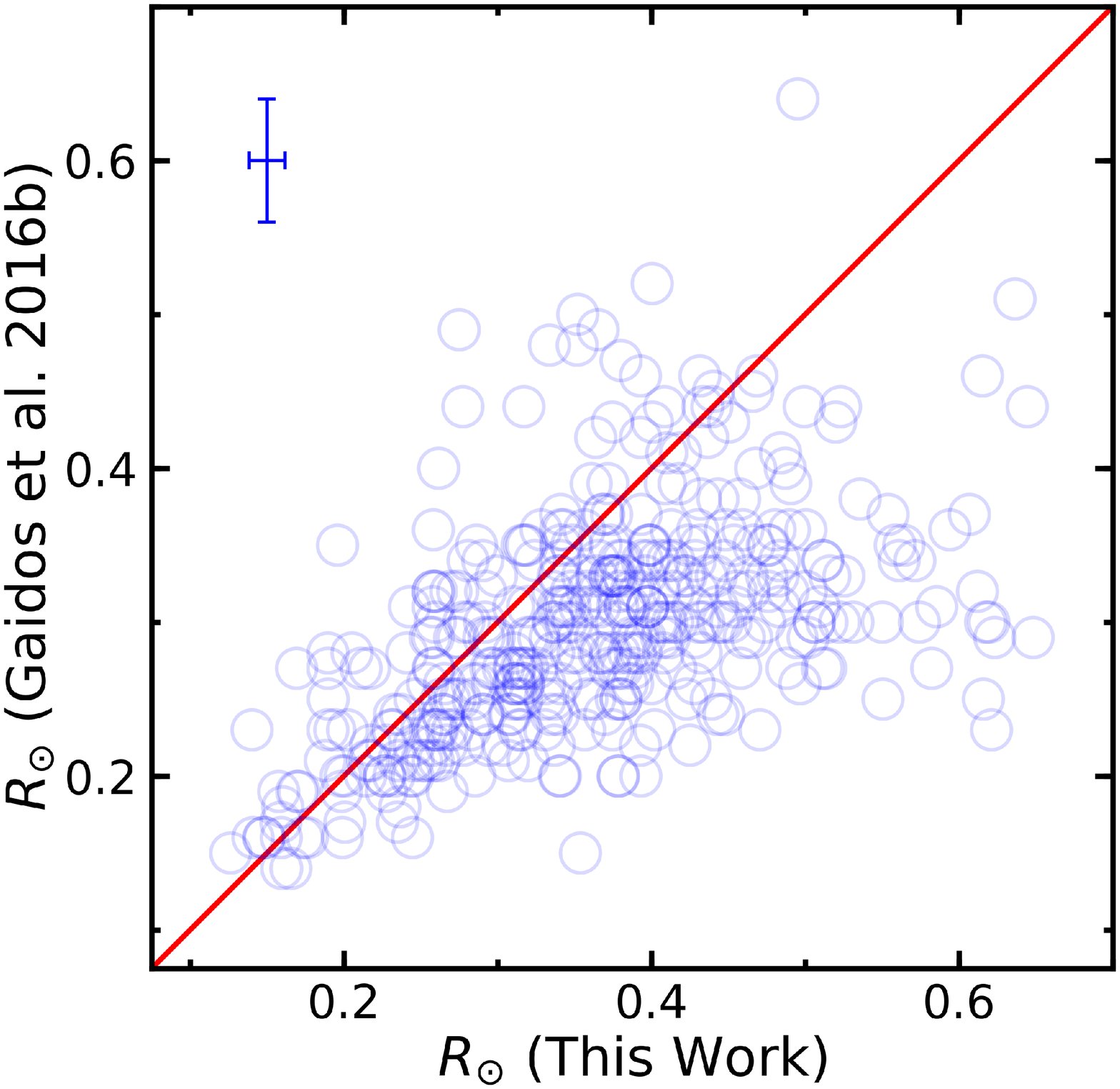}{0.3\textwidth}{\ \ \ \ \ \ \ \ (c)}}
\caption{(a) Comparison of our radius measurements of \Kepler\ M dwarfs to radii from the \Kepler\ Stellar Catalog \citep[][]{Mathur2017}. We do not include 17 targets from the \Kepler\ Stellar Catalog with $R_{\star} > 1\ R_{\odot}$. The 1:1 reference line is plotted in red, and median uncertainties are shown in the upper left corner. The histograms show the radius measurement distributions. (b) Comparison of our radius measurements to those from \citetalias{Dressing2013} for the 213 targets found in both samples. (c) Comparison of our radius measurements to those from \citet{Gaidos2016b} for the 399 targets found in both samples. Our radii are generally larger than previous measurements.} \label{fig:rad}
\end{figure*}

\subsection{Radius}
\label{sec:rstar}

For the 95\% of targets with distance measurements, we apply the $M_{K_s}$ vs.\ $R_{\star}$ and $M_{K_s}$ vs.\ $\textrm{[Fe/H]}$ vs.\ $R_{\star}$ relationships of \citetalias{Mann2015} to compute stellar radii, which have model fit uncertainties of 2.89\% and 2.70\%, respectively. We employ the same MC analysis from Section~\ref{sec:absmag} to assess parameter uncertainties, and add them to the model uncertainties in quadrature, which yields a median 3.13\% and 2.85\% radius uncertainty, respectively.

Not all of our targets have distance measurements, so we rely on the \citetalias{Mann2015} $T_{\mathrm{eff}}$ vs.\ $R_{\star}$ and $T_{\mathrm{eff}}$ vs.\ $\textrm{[Fe/H]}$ vs.\ $R_{\star}$ relationships to get the remaining stellar radii. Adding in quadrature the 13.4\% ($T_{\mathrm{eff}}$ vs.\ $R_{\star}$) and 9.3\% ($T_{\mathrm{eff}}$ vs.\ $\textrm{[Fe/H]}$ vs.\ $R_{\star}$) uncertainties to values from the MC analysis, we yield median uncertainties of 20.6\% and 18.6\%, respectively, for these targets.

In Figure~\ref{fig:rad} we compare our new stellar radius measurements to those reported in the \Kepler\ Stellar Catalog \citep{Mathur2017}, 213 of our targets also found in \citetalias{Dressing2013}, and 399 stars in \citet{Gaidos2016b}. Both the \Kepler\ Stellar Catalog and \citetalias{Dressing2013} show a similar radius distribution for most targets, with a peak near $0.2\,R_{\odot}$, whereas our measurements are generally larger by $\sim$58\%. Our radii are more consistent with those reported in \citet{Gaidos2016b}, but our use of distances from \Gaia\ yield more precise values of $M_{K_s}$, resulting in $\sim$20\% larger radii.

\subsection{Mass}
\label{sec:mstar}

We apply the $M_{K_s}$ vs.\ $M_{\star}$ relationship of \citet{Mann2019} to compute stellar masses of targets with distance measurements using the Python code provided with that paper\footnote{\url{https://github.com/awmann/M_-M_K-}}. This yields median uncertainties of 2.59\% and 2.64\% for targets with and without metallicity measurements, respectively. For the 28 targets without $M_{K_s}$ measurements we need another method to compute mass. We fit the mass and radius measurements for 183 M dwarfs in \citetalias{Mann2015} with a third degree polynomial using the \texttt{polyfit} routine in the \texttt{numpy} \citep{Oliphant2015} Python package and find:
\begin{eqnarray}
\label{eqn:massrad}
    M_{\star}=-0.01+0.63R_{\star}+1.34R_{\star}^2-0.99R_{\star}^3.
\end{eqnarray}
The root-mean square scatter of this fit is 3.75\%. We apply Equation~\ref{eqn:massrad} to the targets for which we do not have distance measurements in the same manner as our previous model fits and yield a median 33.2\% mass uncertainty for these targets.

\section{Refining the Sample}
\label{sec:gb}

Our target selection criteria were chosen to isolate M dwarfs, but that does not preclude a few giant or binary interlopers. From our spectra, we identify four giants (Section~\ref{sec:giants}), two eclipsing binaries, four potential white dwarf-M dwarf binaries (Section~\ref{sec:binaries}), and 15 probable binaries from the \Gaia\ data (Section~\ref{sec:gaiacut}).

\subsection{Giants}
\label{sec:giants}

\begin{figure}[ht!]
    \centering
    \includegraphics[width=0.47\textwidth]{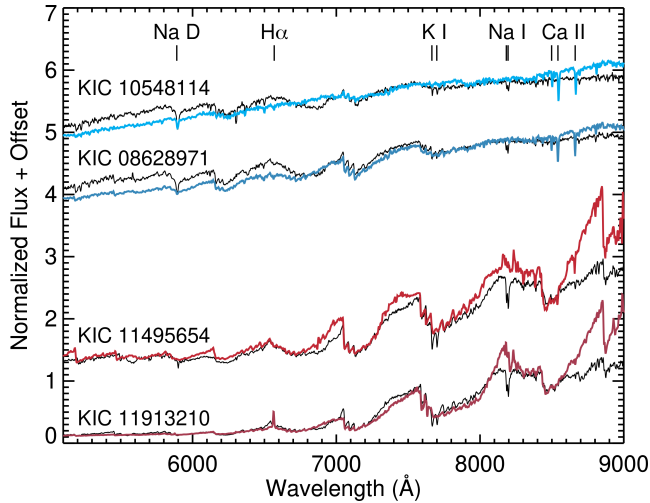}
    \caption{Four giant stars identified in our data (in color), corresponding to two early and two late-type M giants. We plot an M dwarf of similar spectral class (black) to show differences in spectral features. Prominent spectral lines used to delineate dwarf and giant spectra are identified at the top of the plot.}\label{fig:giants}
		\vspace{-0.75em}
\end{figure}

Spectroscopic indicators of M giants include weak Na~D, K~I, and Na~I absorption, significant Ca~II absorption, and differing shapes in the CaH2 index compared to M dwarfs \citep{Reid1995,Mann2012}. The four giants we identify from spectra are KIC~08628971, KIC~10548114, KIC~11495654, and KIC~11913210 (Figure~\ref{fig:giants}). KIC~11913210 is CH Cygni, an S-type symbiotic star system\footnote{An S-type star is a cool giant containing zirconium monoxide bands in their spectra \citep{Merrill1922,Merrill1923}. This is not to be confused with an S-type orbit, where a planet orbits a single star in a binary star system.} consisting of at least two giant stars \citep{Skopal2005}. We discard these sources in our occurrence rate analysis, reducing our sample to 557 stars.

\subsection{Binaries}
\label{sec:binaries}
Within the \Kepler\ data, 2,909 eclipsing binaries have been identified\footnote{\url{http://keplerebs.villanova.edu/}, as of April 2019}, which is $\sim$1.5\% of the nearly 200,000 \Kepler\ stars observed \citep{Kirk2016,Abdul-Masih2016}. Two of our targets are eclipsing binaries, KIC~07174349 and KIC~10002261, which we spectral type as M4\,V and M3.5\,V, respectively. 

\begin{figure}[ht!]
    \centering
    \includegraphics[width=0.47\textwidth]{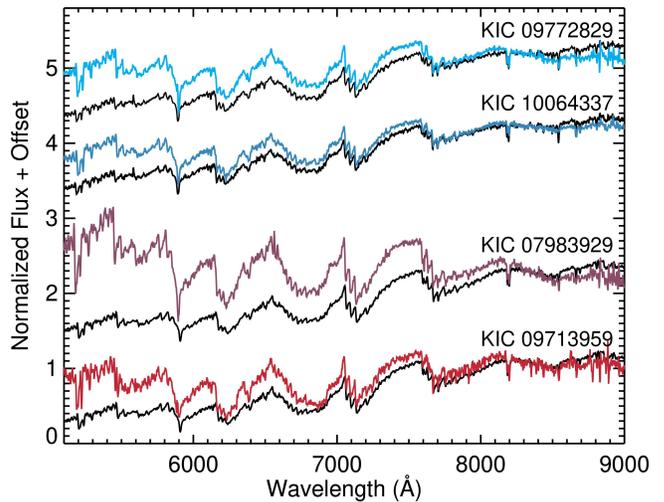}
    \caption{Four potential white dwarf-M dwarf binaries. KIC~09772829 and KIC~100064337 are compared an M2\,V spectrum, and KIC~07983929 and KIC~09713959 are compared to an M3\,V spectrum (black). Spectra are normalized to 8,350$\,$\AA.}\label{fig:mwd}
		\vspace{-0.75em}
\end{figure}

Our spectra are not high enough resolution to be able to detect single- and double-lined spectroscopic binaries, however, we identify four targets which are potential white dwarf-M dwarf binaries. The spectra of KIC~07983929, KIC~09713959, KIC~09772829, and KIC~10064337 have excess flux toward blue wavelengths (Figure~\ref{fig:mwd}), which we attribute to a companion. Additionally, these spectra have similar spectral shapes to red-optical spectra of white dwarf--M dwarf binaries in the literature \citep[e.g.,][]{Ren2014,Skinner2017}. It is possible that the excess blue flux could be due to either improper flux calibration, slit losses, or atmospheric dispersion. Since all these binary candidates were observed using Hydra, we checked the spectra of other M dwarfs observed at the same time in the same field configuration and find no excess blue flux in these targets. Additionally, none of these targets were observed with the same fiber, and other targets observed with the same fiber in a different fiber configuration do not exhibit excess blue flux. We recommend additional spectroscopic observations of these targets at red and blue wavelengths to check that these effects are not from instrument systematics, and to confirm our assessment that these systems are likely white dwarf-M dwarf binaries. \citet{Wang2014} and \citet{Kraus2016} find that close binary companions appear to suppress planet formation, and hence decrease planet occurrence rates for these systems. As such, we remove the eclipsing and candidate white dwarf-M dwarf binaries from our analysis of planet occurrence rates, reducing our sample to 551 stars.

\subsection{Gaia Color Cuts and Binaries}
\label{sec:gaiacut}
We identify 61 targets with more than one source within 4\arcsec\ of the KIC coordinates, in which case we choose the target with the smallest difference in $G-Kp$ magnitudes. We also check if the $M_{K_s}$ magnitudes (Section~\ref{sec:absmag}) are in the range we expect for an M dwarf ($4.6 \lesssim M_{K_s} \lesssim 9.8$; \citetalias{Mann2015}). Four targets for which we do not have spectra (KIC~10711066, KIC~10473048, KIC~06757650, and KIC~06029053) have \Gaia\ coordinates $< 0\farcs4$ from the KIC coordinates, but have $M_{K_s}$ magnitudes of 2.0, -1.6, -3.7, and 3.2, respectively. We discard these four sources from our analysis as they are likely non-M dwarfs, reducing our sample to 547 stars. Two spectroscopically observed targets, KIC~04545041 (M3.5\,V) and KIC~03631048 (M4\,V) have $M_{K_s}$ of 4.06 and 4.45, respectively. Additionally, two photometrically classified targets, KIC~11196417 (M5\,V) and KIC~10991155 (M4\,V), have $M_{K_s}$ of 4.53 and 4.56, respectively. For these four targets, we use their effective temperatures to determine radius since the $M_{K_s}$ vs.\ $R_{\star}$ relationships do not extend to $M_{K_s} < 4.6$.

\begin{figure*}[ht!]
    \centering
    \includegraphics[width=0.99\textwidth]{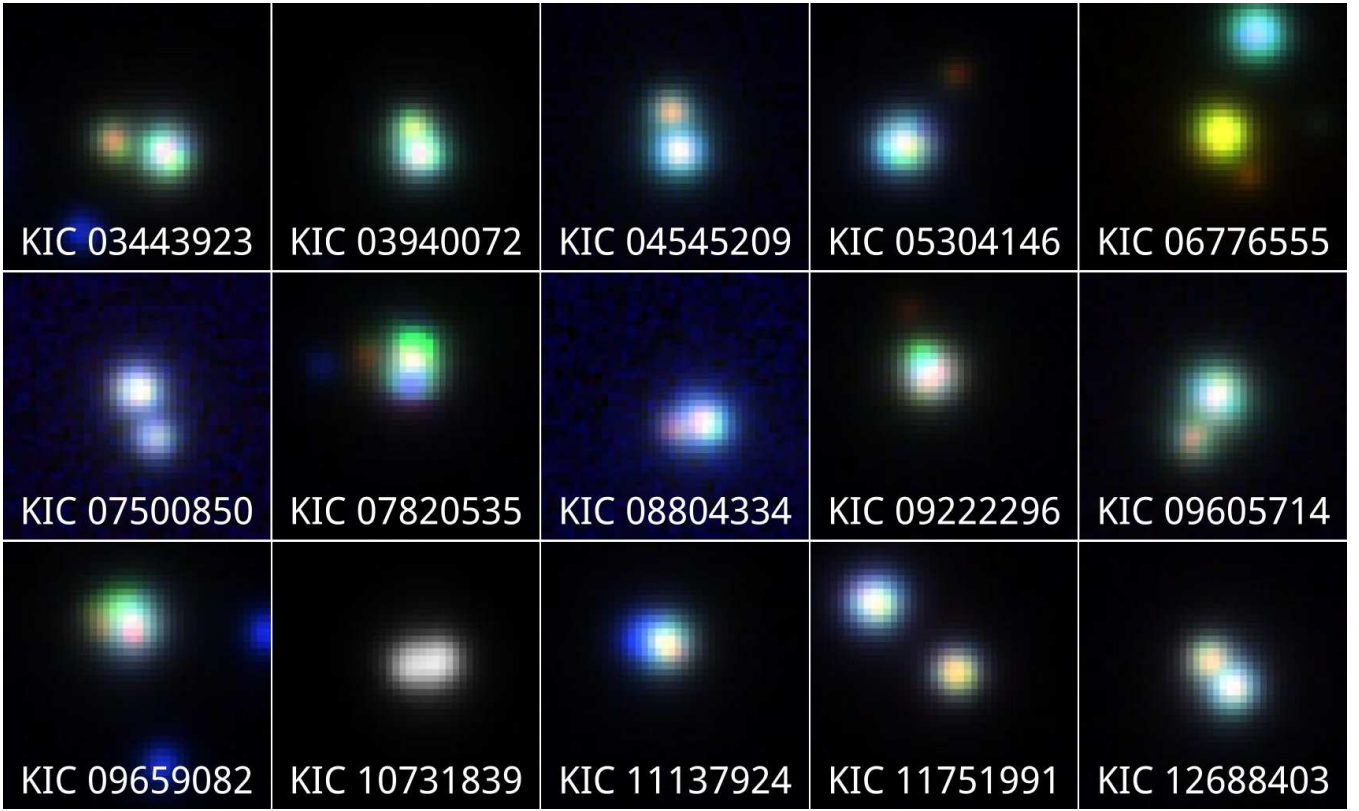}
    \caption{Probable binary stars sharing common proper motions and distances in the \Gaia\ data. These 10\arcsec\ cutouts are stacked $g$, $i$, and $y$-band images from Pan-STARRS and are centered on the target coordinates from the NASA Exoplanet Archive. The image for KIC~10731839 is only $g$-band since it is saturated in the other filters.}\label{fig:visb}
	\vspace{0em}
\end{figure*}

Within the \Gaia\ data, 15 targets have a nearby star with a similar distance and common proper motion, which we identify as probable binaries. In Figure~\ref{fig:visb} we show 10\arcsec\ cutout images of these 15 targets from the Panoramic Survey Telescope and Rapid Response System survey \citep[Pan-STARRS,][]{Chambers2016}. We do not discard these stars in our analysis, since we use the $G-Kp$ color selection to identify our primary targets. In total, we have identified 21 binaries and binary candidates, which is 3.65\% of our total sample. \citet{Duchene2013} report a decreasing stellar multiplicity frequency with decreasing stellar mass, and estimate that $26\pm3\%$ of M dwarfs are in multiple systems. \citet{Winters2019} find an M dwarf multiplicity of $26.8\pm1.4\%$, and also note that lower mass M dwarfs have fewer companions than higher mass M dwarfs. We recognize that there are likely more unresolved binaries in our sample since we detect a $\sim$4\% binary fraction, much lower than the expected $\sim$27\%, however, the assessment of the effect of binaries on planet occurrence rates is beyond the scope of this paper.

\section{Mid-Type M Dwarf Planets}
\label{sec:planets}

\begin{figure*}[ht!]
\gridline{\fig{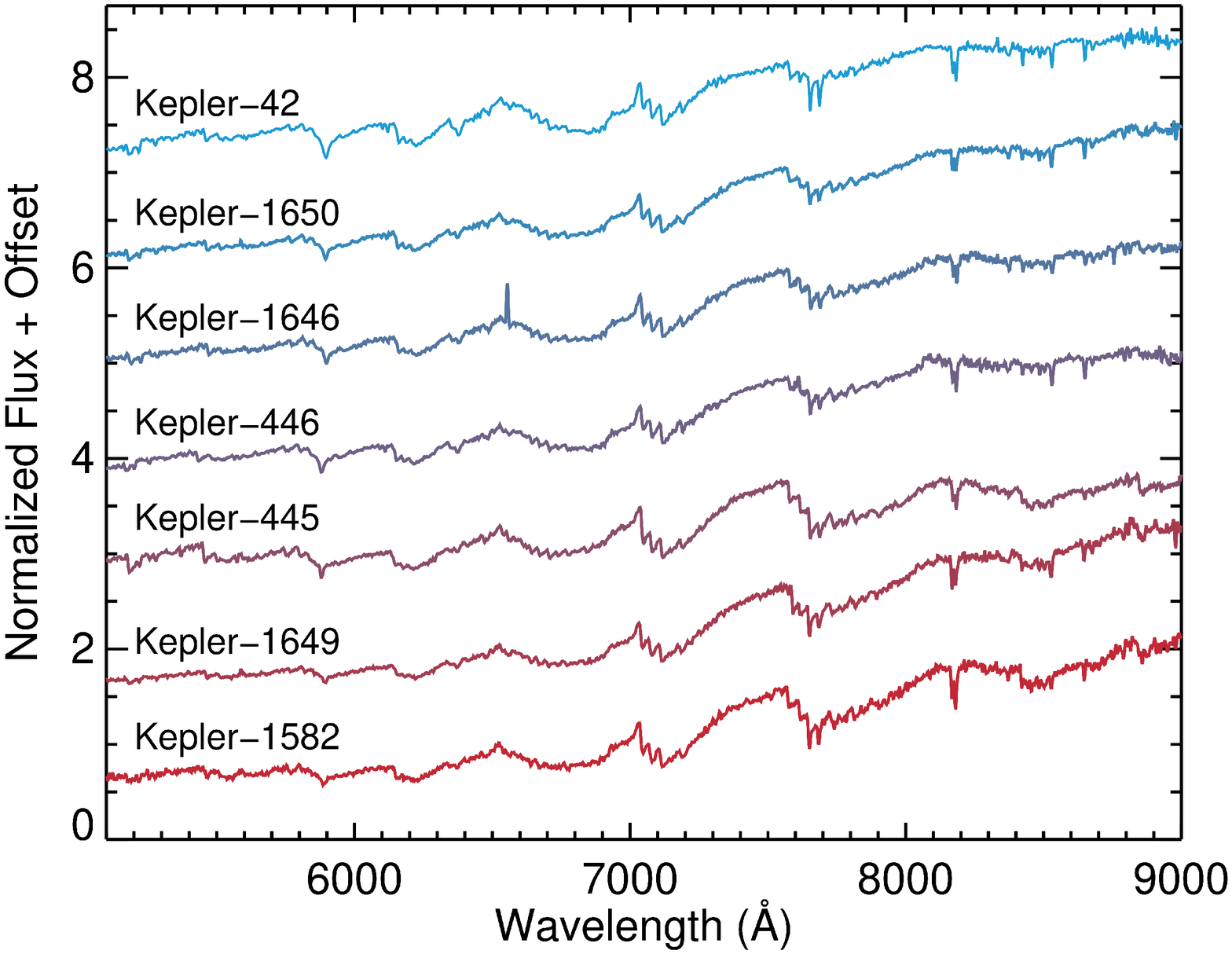}{0.46\textwidth}{\ \ \ \ \ \ \ \ \ \ (a)}
          \fig{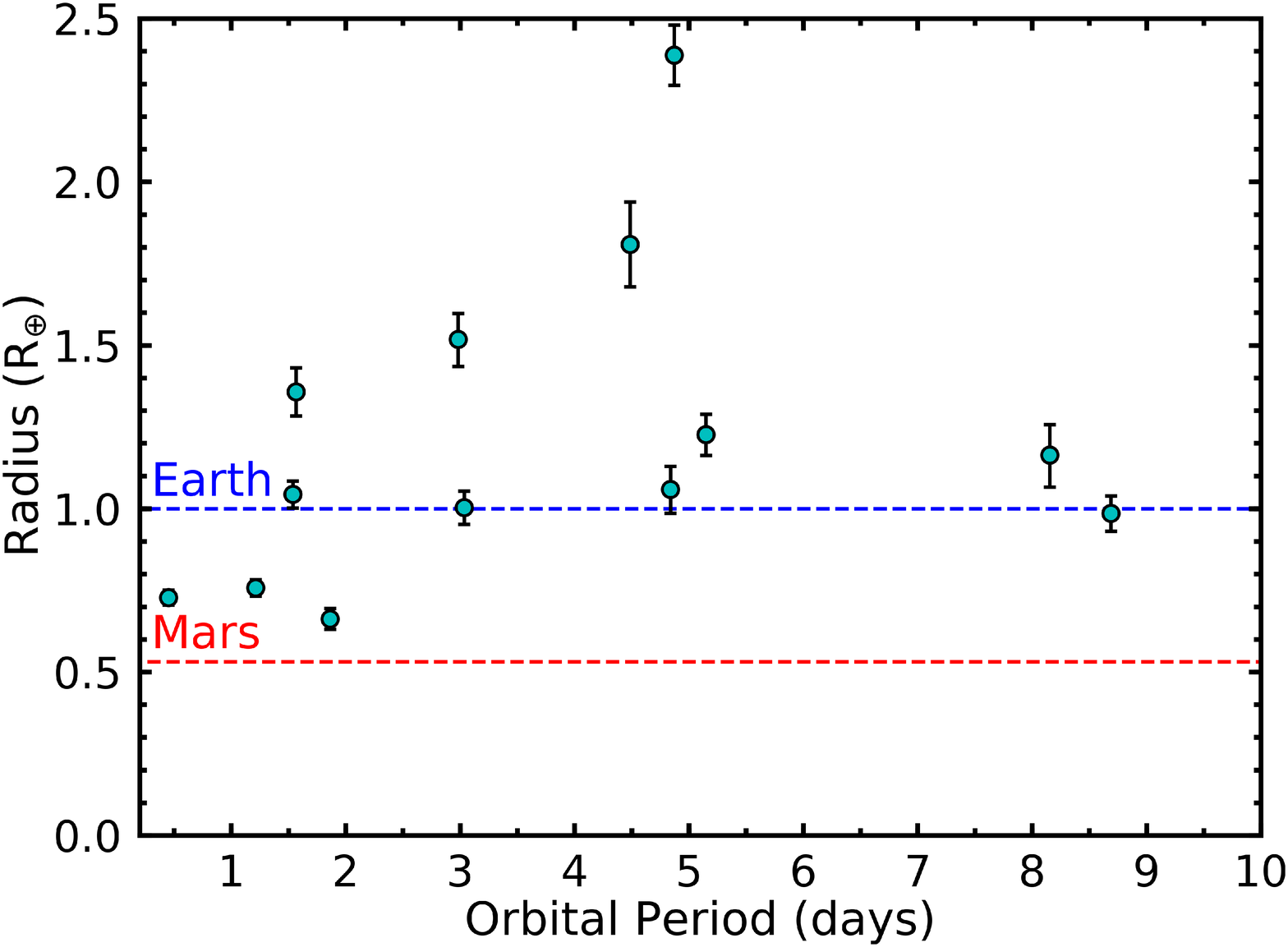}{0.47\textwidth}{\ \ \ \ \ \ \ \ \ (b)}}
\caption{(a) Planet host spectra for \Kepler\ mid-type M dwarfs, ordered by spectral type from M3\,V (top) to M5\,V (bottom). These are the first published optical spectra for Kepler-1646 (M4\,Ve) and Kepler-1582 (M5\,V). (b) Orbital period versus planet radius for \Kepler\ mid-type M dwarfs. The planet radii have been updated using our new stellar radius measurements. The radii of Earth and Mars are shown for reference.}\label{fig:radper}
\end{figure*}

With our updated stellar properties in hand, we now turn to planets around mid-type M dwarfs. In our \Kepler\ star sample, there are six planet systems containing 12 confirmed planets. We also include Kepler-1650~b, for which the host star has an $r-J$ color of 3.167, just missing our color cut. \citet{Muirhead2014} identify Kepler-1650 as an M3\,V star from $H$ and $K$-band spectra. Our optical spectrum of this target gives us a classification of M3.5\,V. Spectra for all planet host stars are shown in Figure~\ref{fig:radper}.

Additionally, there are four Kepler Objects of Interest (KOI) within our sample: KOI-959, KOI-5237, KOI-6705, and KOI-8012. Based on our spectra, KOI-959 (KIC~10002261), KOI-5327 (KIC~06776555), KOI-8012 (KIC~10452252) have spectral types of M3.5\,V, M4\,V, and M5\,V, respectively. Using the spectral index typing system of \citet{Lepine2013}, \citet{Gaidos2016} report KOI-6705 (KIC~06423922) to be an M3.4\,V star from a SNIFS optical spectrum.

\citet{Slawson2011} identify KOI-959 to be an eclipsing binary, which makes this planet candidate an astrophysical false positive. Additionally, \citet{Furlan2017} and \citet{Ziegler2018} find a nearby companion to KOI-959 at a separation of $\sim$0$\farcs77$ using adaptive optics imaging. KOI-5327 is associated with three nearby companions at separations of $1\farcs88$, $3\farcs63$, and $3\farcs96$ \citep{Ziegler2017,Furlan2017}. The \Kepler\ pipeline data validation report for this system shows an out-of-transit centroid offset of $2\farcs09$ (6.3$\sigma$), which might be due to the closest companion star. It is unclear which star the planet orbits, so we do not include this system in our analysis. \citet{Gaidos2016} identify KOI-6705.01 as a likely false positive due to charge transfer inefficiency in a detector column on which a nearby eclipsing binary falls. \citet{Thompson2018} identify KOI-8012.01 as a Mercury-sized planet candidate in the habitable zone of its host star. Our reassessment of the radius of KOI-8012 places it at $0.43\,R_{\odot}$, which in turn increases the planet candidate radius to $0.83\,R_{\oplus}$. We do not include this planet candidate in our analysis since the system parameters (e.g., period, semi-major axis) do not have reported uncertainties, which makes accurate analysis of this system difficult. Additionally, the \Kepler\ light curve for KOI-8012 shows significant variability and extreme flare activity, which suggests that any planet candidates around this star could be astrophysical false positives.

Table~\ref{tab:planets} and Figure~\ref{fig:radper} show the confirmed planets around mid-type M dwarfs in the \Kepler\ field. We note that all these planets are smaller than Neptune ($3.865\,R_{\oplus}$) and have orbital periods shorter than 10 days. We update the planet parameters for all these systems based on our stellar radius and mass measurements, and compute insolation flux $S_p$ using:
\begin{equation}
    \Bigg(\frac{S_p}{S_{\oplus}}\Bigg)=\Bigg(\frac{L_{\star}}{L_{\odot}}\Bigg)\Bigg(\frac{AU}{a}\Bigg)^2,
\end{equation}
where $(L_{\star}/L_{\odot})=(R_{\star}/R_{\odot})^2 (T_{\mathrm{eff}}/T_{\odot})^4$ is the stellar luminosity, and $a$ is the semi-major axis in AU, computed from Kepler's third law using planet orbital period from the literature and our stellar masses. The planet with the lowest stellar irradiance in our sample is Kepler-1649~b, an Earth-sized planet ($R_p=0.985\pm0.054\,R_\oplus$) receiving nearly twice the stellar irradiation as Earth. The habitable zone definitions of \citet{Kopparapu2013} place this planet above the maximum greenhouse stellar irradiance limit.

\section{Planet Occurrence Rate}
\label{sec:occur}
With our updated stellar properties of 547 stars and a population of 13 planets around mid-type M dwarfs, we now have the critical pieces necessary to compute planet occurrence rates. We adopt the grid-based method \citep[e.g,][]{Howard2012, Dressing2013, Petigura2013} to compute planet occurrence rates, $f(R_{p},P)$, which is the fraction of stars in a population (e.g., mid-type M dwarfs) that have planets with specific radii $R_p$ and orbital periods $P$. Using the population of known planets around \Kepler\ mid-type M dwarfs, we count the number of \Kepler\ mid-type M dwarfs $N_{\star,i}$ around which the $i^{th}$ known planet could have been detected with sufficient photometric precision if it was at a transiting orbital inclination:
\begin{equation}
\label{eq:occurrence}
    f(R_{p},P)=\sum_{i=1}^{N_p}\frac{1/p_i}{N_{\star,i}},
\end{equation}
where the sum is over the number of known planets $N_p$ from the \Kepler\ data within the specified planet radius and orbital period range. The probability $p$ that planet $i$ is in a transiting geometry, is defined as:
\begin{equation}
    p_i=\Bigg(\frac{R_{\star,i}+R_p}{a_i}\Bigg)\Bigg(\frac{1+e_i \sin \omega_i}{1-e_{i}^{2}}\Bigg),
\end{equation}
where $a_i$ is the semi-major axis of planet $i$, $e_i$ is the eccentricity, and $\omega_i$ is the longitude of periastron \citep{Winn2010,Stevens2013}. Typically, $R_p \ll R_{\star}$ and small planets on close-in orbits are nearly circular, which makes $p_i \approx R_{\star,i}/a_i$. However, an Earth-sized planet around a mid-type M dwarf is between 2.5 and 7.5\% of its host star diameter, which can have a small effect on the transit probability. Assuming circular orbits, we use:
\begin{equation}
\label{eq:transprob}
    p_i=\Bigg(\frac{R_{\star,i}+R_p}{a_i}\Bigg).
\end{equation}

We compute $N_{\star,i}$ from the number of stars for which the S/N from the \Kepler\ photometry is high enough that a known planet could be detected. The S/N is calculated by:
\begin{equation}
\label{eq:sn}
	\text{S/N} = \frac{R^2_{\text{p}}/R^2_{\star}}{\text{CDPP}_{\text{eff}}}
	\sqrt{\frac{t_{\text{obs}}}{P}},
\end{equation}
where $\text{CDPP}_{\text{eff}}$ is the Combined Differential Photometric Precision over the duration of the transit, $t_{\text{obs}}$ is the total time spent observing the star with \Kepler, and $P$ is the known planet orbital period \citep{Muirhead2015}. $\text{CDPP}_{\text{eff}}$, more precisely, is an estimate of white noise during a specific transit duration \citep{Christiansen2012}. For each star, we fit a power law to the CDPP values versus time for transits between 1.5 and 15 hours, and use this fit to determine $\text{CDPP}_{\text{eff}}$ for each planet transit duration, $t_{\mathrm{dur}}$. We compute $t_{\mathrm{dur}}$ for known planets using:
\begin{equation}
\label{eq:tdur}
    t_{\mathrm{dur}}=\frac{P}{\pi}\sin^{-1}\Bigg(\frac{\sqrt{(R_{\star}+R_p)^2-(bR_{\star})^2}}{a}\Bigg),
\end{equation}
where $b=a\cos \theta /R_{\star}$ is the impact parameter with orbital inclination angle $\theta$.

We use the linear ramp detection efficiency model from \citet{Mulders2015}, which was adapted from \citet{Fressin2013}, where detection efficiency $f_\mathrm{eff}=0$ for S/N~$\leq$ 6, $f_\mathrm{eff}=1$ for S/N $>$ 12, and $f_\mathrm{eff}=(\mathrm{S/N}-6)/6$ for 6 $<$ S/N $\leq$ 12. The sum of the detection efficiencies, rounded to the nearest integer, then becomes $N_{\star,i}$.

Our stellar selection criteria were chosen to isolate stars with a spectral type of M3\,V and beyond, though we observed 40 stars with earlier spectral types. In our planet sample, we have identified the \Kepler\ planets around M3\,V to M5\,V stars. For our planet occurrence rate calculations, we only consider the stars in this spectral type range.

\begin{figure*}[ht!]
    \centering
    \includegraphics[width=0.99\textwidth]{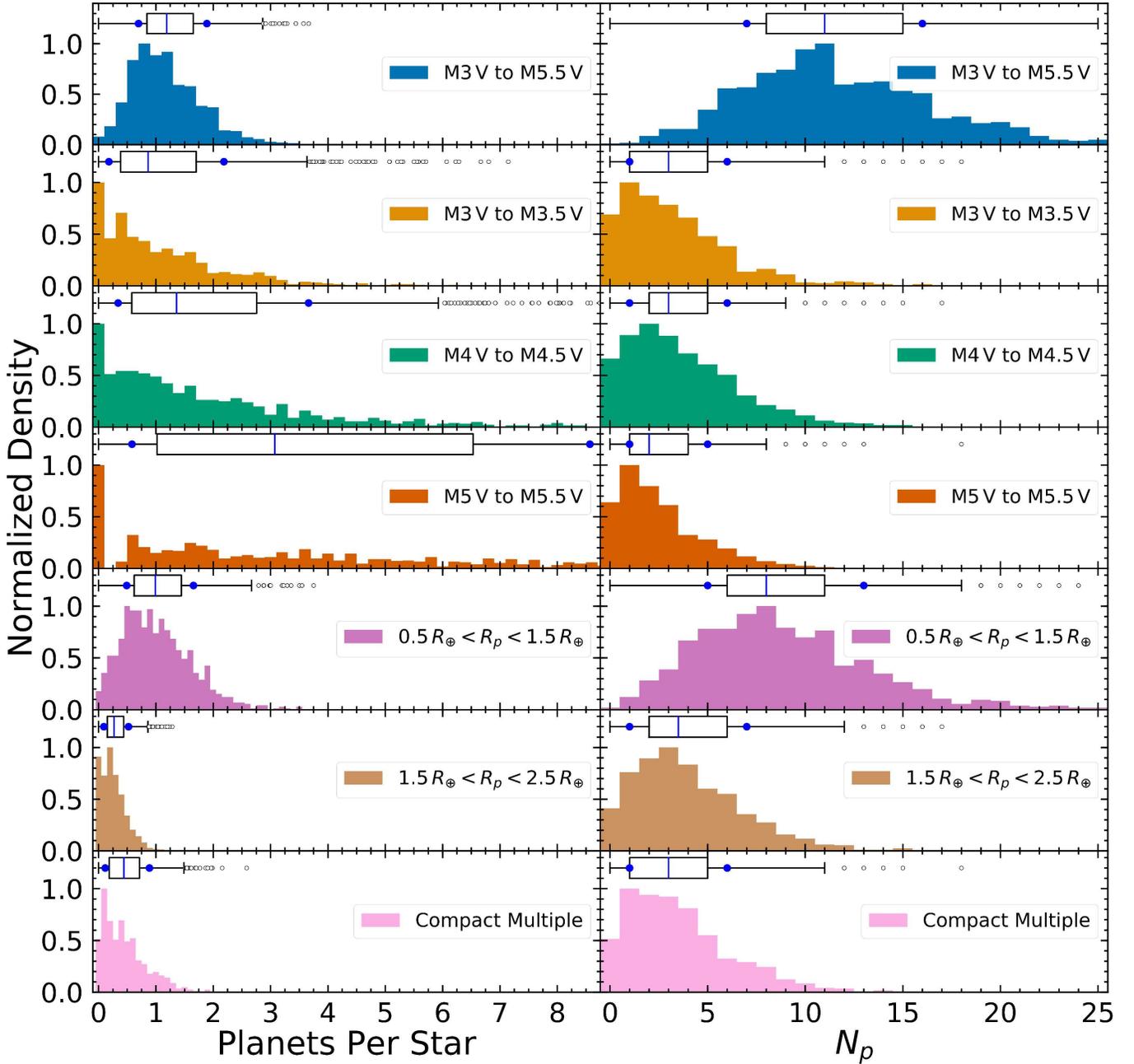}
    \caption{(Left) Planet occurrence rate distributions from our Monte Carlo analysis for different spectral type bins, planet radius bins, and for compact multiple systems. Above each distribution we give the corresponding box-and-whisker plot, which show the quartiles of each distribution. The 16$^{\mathrm{th}}$ and 84$^{\mathrm{th}}$ percentiles are shown as blue circles, and any outliers beyond $Q3+1.5\times IQR$ are shown as open circles. The M5\,V upper whisker extends to $\sim$15, however, we restrict the horizontal range to just beyond the 84$^{\mathrm{th}}$ percentile of the M5\,V distribution for clarity in the other plots. All of these plots are normalized such that the peak of each distribution is equal to one. (Right) Planet detection distributions from the planet occurrence rate Monte Carlo analysis and corresponding box-and-whisker plots.}\label{fig:pormc}
		\vspace{0em}
\end{figure*}

\subsection{Monte Carlo Analysis}
\label{sec:MC}
The measured star and planet parameters found in Equations~\ref{eq:transprob}, \ref{eq:sn}, and \ref{eq:tdur} have uncertainties which contribute to the total uncertainty of the planet occurrence rate calculation. There is also spectral type uncertainty, which can move stars, including planet hosts, into or out of a spectral type bin. Additionally, due to our small planet sample size, we need to account for counting errors. In order to address these uncertainties we perform the following MC analysis. First, we address the $\pm1$ spectral type uncertainty by drawing a random number from a uniform distribution between -1 and 1 for each star in our sample and add those numbers to each spectral type. Second, we draw a number of planet detections $N_p$ from a Poisson distribution with an expectation value ($\lambda$) equal to the detected planet sample size, similar to the method used in \citet{Silburt2015} and \citet{Gaidos2016b}. For example, our total \Kepler\ mid-type M dwarf planet sample size is 13, so we randomly draw a number from a Poisson distribution with $\lambda=13$, then sample with replacement that number of planets from our mid-type M dwarf planet catalog to run the occurrence rate calculations. Next, for each parameter with an uncertainty measurement in Equations~\ref{eq:transprob}, \ref{eq:sn}, and \ref{eq:tdur}, we generate $10^4$ random samples from a Gaussian distribution and perform each calculation with these distributions. We also account for the effect of transit duration on S/N by assigning a random inclination angle to compute the impact parameter in Equation~\ref{eq:tdur}. We then run the results through Equation~\ref{eq:occurrence}, and repeat this procedure $10^3$ times for seven different bins: all mid-type M dwarfs (M3\,V to M5.5\,V), individual spectral types (M3\,V to M3.5\,V, M4\,V to M4.5\,V, and M5\,V to M5.5\,V), Earth-sized planets ($0.5\,R_{\oplus} < R_p < 1.5\,R_{\oplus}$), super-Earths ($1.5\,R_{\oplus} < R_p < 2.5\,R_{\oplus}$), and compact multiples.

The resulting planet occurrence rates and planet detection distributions are shown in Figure~\ref{fig:pormc}. The occurrence rate distributions all exhibit positive skew, highlighted by the box-and-whisker plots above each distribution. Due to our random selection of $N_p$ from a Poisson distribution, an outcome of zero planets is possible, which results in a planet occurrence rate of zero. For our seven different bins, those with a smaller initial sample of planets are more likely to randomly draw zero planets. This effect is most prominent for the individual spectral type bins, which have prominent peaks at zero for the planet occurrence rate. Even though the $N_p$ distributions have a non-zero median, drawing zero planets will always result in an occurrence rate of zero, whereas drawing one or more planets can result in a different planet occurrence rate depending on the transit probability of the randomly selected planets and number of stars from the MC calculation, effectively smoothing out other peaks. We report the results from the planet occurrence rate calculation in Table~\ref{tab:subpor}.

\begin{deluxetable}{ccll}[htbp]
\tablecaption{Planet occurrence rates per spectral type, planet radius, and compact multiples for mid-type M dwarfs.\label{tab:subpor}}
\tablewidth{0pt}
\tablehead{
\colhead{Range} & \colhead{Planets per star} & \colhead{$N_p$} & \colhead{$N_{\star,i}$}
}
\startdata
M3\,V to M5.5\,V & $1.19^{+0.70}_{-0.49}$ & $11^{+5}_{-4}$ & $412^{+8}_{-8}$ \\
\hline
M3\,V to M3.5\,V & $0.86^{+1.32}_{-0.68}$ & $3^{+3}_{-2}$ & $95^{+8}_{-8}$ \\
M4\,V to M4.5\,V & $1.36^{+2.30}_{-1.02}$ & $3^{+3}_{-2}$ & $92^{+9}_{-8}$ \\
M5\,V to M5.5\,V & $3.07^{+5.49}_{-2.49}$ & $2^{+3}_{-1}$ & $37^{+5}_{-6}$ \\
\hline
$0.5\,R_{\oplus} < R_p < 1.5\,R_{\oplus}$ & $0.99^{+0.66}_{-0.50}$ & $8^{+5}_{-3}$ & $412^{+8}_{-9}$ \\
$1.5\,R_{\oplus} < R_p < 2.5\,R_{\oplus}$ & $0.27^{+0.25}_{-0.18}$ & $4^{+3}_{-3}$ & $412^{+7}_{-9}$ \\
\hline
Compact Multiple & $0.44^{+0.45}_{-0.33}$ & $3^{+3}_{-2}$ & $411^{+8}_{-8}$ \\
\enddata
\tablecomments{We report the median, 16$^{\mathrm{th}}$, and 84$^{\mathrm{th}}$ percentiles of each distribution from $10^3$ Monte Carlo trials.}
\vspace{-18pt}
\end{deluxetable}

\section{Discussion}
\label{sec:discussion}
\begin{figure*}[ht!]
    \centering
    \includegraphics[width=0.99\textwidth]{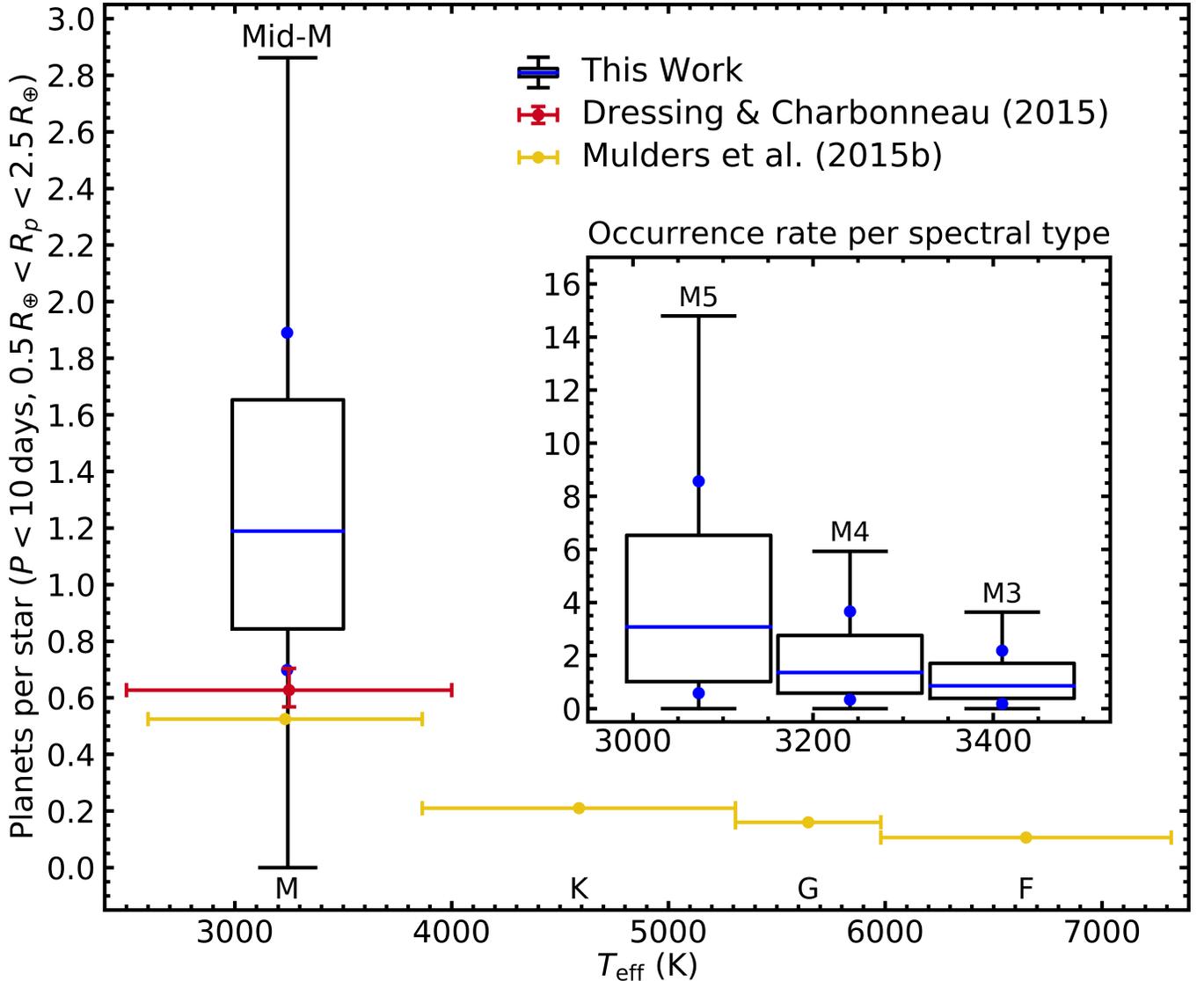}
    \caption{Planet occurrence rate as a function of stellar effective temperature for planets with orbital periods shorter than 10 days and radii between 0.5 and $2.5\,R_{\oplus}$. The measurements from our Monte Carlo analysis are shown as box-and-whisker plots (see Figure~\ref{fig:pormc}), where the blue line indicates the median value, the blue dots are the 16$^{\mathrm{th}}$ and 84$^{\mathrm{th}}$ percentiles of the distributions, and the box width corresponds to the temperature uncertainty. The inset plot shows the planet occurrence rates for M3\,V, M4\,V, and M5\,V, which show evidence for an increase of planet occurrence toward later M dwarfs.}\label{fig:por}  
		\vspace{0em}
\end{figure*}

For early-type M dwarfs with orbital periods between 0.5 and 10 days and planets with $0.5\,R_{\oplus} < R_p < 2.5\,R_{\oplus}$, \citetalias{Dressing2015} derive a planet occurrence rate of $0.63^{+0.08}_{-0.06}$ planets per star. Similarly, \citet{Mulders2015b} compute a planet occurrence rate for M dwarfs in this range to be 0.53 planets per star. Our total occurrence rate for mid-type M dwarfs is approximately double that of \citetalias{Dressing2015} and \citet{Mulders2015b} for early-type M dwarfs. We show a comparison of occurrence rates from our MC analysis to values from \citetalias{Dressing2015} and \citet{Mulders2015b} in Figure~\ref{fig:por}. We also include in this figure the planet occurrence rates for stars of each spectral type F, G, and K from \citet{Mulders2015b}. There is a clear trend of increasing planet occurrence toward later spectral types. If we take the median values of planet occurrence rates from our MC analysis, this suggests an increasing trend of small planets at short orbital periods toward later-type M dwarfs. We also find that a typical mid-type M dwarf will host a short period Earth-sized planet ($0.5\,R_{\oplus} < R_p < 1.5\,R_{\oplus}$), and larger planets are about four times less common.

We revise the planet occurrence rate for compact multiples around mid-type M dwarfs to be $0.44^{+0.45}_{-0.33}$. This value is computed following the same MC procedure, but only considers the detectability of the outermost planet in the multi-planet system. Our occurrence rate for compact multiples is consistent within $1\sigma$ to $0.21^{+0.07}_{-0.05}$ reported by \citet{Muirhead2015}, though our median value is higher due to our revised stellar radii. \citet{Muirhead2015} assumed a uniform stellar radius of 0.2~$R_{\odot}$, the peak of the radius distribution for these stars in the \Kepler\ Stellar Catalog. For low mass stars, \citet{Boyajian2012} found that evolutionary models, which were largely used to determine radii in the \Kepler\ Stellar Catalog, under predict stellar radii. Our radius measurements have a median value of $0.33\,R_{\odot}$, 58\% larger than the median $0.21\,R_{\odot}$ from the \Kepler\ Stellar Catalog. The stellar radii reported in \citet{Gaidos2016b} were a significant improvement over those derived from evolutionary models, but they are still $\sim$20\% smaller than our measurements (Figure~\ref{fig:rad}).

\citet{Ballard2016} assessed planet multiplicity among \Kepler\ M dwarfs, and found that half of these systems contain five or more co-planar planets, while the other half are either single or multiple planet systems with large mutual planet inclinations. Of the seven confirmed \Kepler\ mid-type M dwarf planet systems, three of them contain at least three planets, whereas the other four systems contain at least one planet. It is possible that all of these systems could contain additional non-transiting planets or planets that fall below the detection threshold of \Kepler. The discovery of seven Earth-sized planets around the M8\,V star TRAPPIST-1 shows that planet formation occurs through the end of the main sequence \citep{Gillon2017}, and supports both the increase in planet occurrence rates toward later spectral types and the large planet multiplicity from \citet{Ballard2016}. 

\citet{Thompson2018} note that pipeline detection efficiency, astrophysical reliability, and imperfect stellar information must all be taken into account for planet occurrence rate calculations. Pipeline detection efficiency has been computed for \Kepler\ F, G, and K stars using pixel-level and flux-level transit injection \citep{Christiansen2016,Christiansen2017,Burke2017}. \citetalias{Dressing2015} performed their own pipeline detection efficiency for \Kepler\ M dwarfs by injecting transit signals into light curves for all their target stars. For planets between $1\,R_{\oplus}$ and $3\,R_{\oplus}$ at orbital periods less than 10 days, their pipeline was able to recover between 80 and 90\% of the injected transit signals. For planets below $1\,R_{\oplus}$, recovery averaged 58\%. Astrophysical reliability tests are used to determine whether or not an observed event is due to instrumental or stellar noise or other astrophysical events. Certain signals can mimic a planet transit (e.g., eclipsing binaries, background eclipsing binaries), so tools have been developed to automatically vet planet candidates \citep[e.g., Robovetter,][]{Thompson2018} and statistically validate planet candidates \citep[e.g., \texttt{vespa},][]{Morton2016}. Our work has addressed the third issue of imperfect stellar information, though we do take into account a simple linear ramp model of detection efficiency in our calculations. We also note that \citet{Silburt2015} and \citet{Gaidos2016b} caution placing planets into discrete radius and period bins because it ignores information about the underlying planet distribution and will underestimate the planet occurrence rate. With a sample of only 13 planets, however, it is difficult to determine a practical planet distribution without making some assumptions. Additionally, accounting for small planet sample size dominates the error budget for our occurrence rate calculations.

From our \Kepler\ planet occurrence rates, we can predict the number of planets in orbital periods less than 10 days around mid-type M dwarfs in the local neighborhood. \citet{Stelzer2013} conducted a survey of UV and X-ray activity of M dwarfs within 10 parsecs of the Sun, in which they observed 159 M0\,V through M8\,V stars, and estimated that their volume-limited M dwarf sample was 90\% complete. This sample includes 101 stars with spectral types between M3\,V and M5.5\,V, which means there are potentially $120^{+71}_{-49}$ small, short period planets around these nearby stars. So far, there have been 20 planets found around 9 of these nearby mid-type M dwarfs, all discovered via the radial velocity technique\footnote{\url{https://exoplanetarchive.ipac.caltech.edu/cgi-bin/TblView/nph-tblView?app=ExoTbls&config=planets}, as of April 2019}. Of those 20 planets, 10 have orbital periods shorter than 10 days. The faintest mid-type M dwarf around which a planet has been detected using the radial velocity technique has a $V$-band magnitude of 12.22 (GJ~3323), and the smallest radial velocity amplitude for a planet detected around these stars is $1.06\pm0.15\,\mathrm{m\,s}^{-1}$ \citep[GJ~273~c,][]{Astudillo-Defru2017}. There are 51 nearby mid-type M dwarfs with $V$-band magnitudes less than 12.22. Our planet occurrence rates suggest there are $61^{+35}_{-25}$ small, short period planets around these stars. Since 10 of these planets have already been found, there may be around 50 additional planets orbiting nearby mid-type M dwarfs which could be detected with current radial velocity instruments \citep[e.g., HARPS,][]{Pepe2011}, assuming the planets are massive enough to induce a $1\,\mathrm{m\,s}^{-1}$ signal. NEID and EXPRES aim to achieve even greater precision below $30\,\mathrm{cm\,s}^{-1}$ at optical wavelengths \citep{Halverson2016,Jurgenson2016}. The Habitable Zone Planet Finder \citep[HPF;][]{Mahadevan2012} will specifically target mid- to late-type M dwarfs at near-infrared wavelengths, with a goal of $1\,\mathrm{m\,s}^{-1}$ precision. Recent commissioning observations with HPF have already achieved $1.3\,\mathrm{m\,s}^{-1}$ precision on Barnard's Star \citep{Mahadevan2018}.

Since the original \Kepler\ mission was designed to detect Earth-like planets around Sun-like stars, less emphasis was placed on low-mass stars. The \textit{K2} mission \citep{Howell2014}, however, was well suited for detecting planets around the latest type stars along the ecliptic plane, and has already produced dozens of confirmed and candidate planets around M dwarfs, opening up another rich data set to expand this planet occurrence rate study. The \textit{K2} M Dwarf Project (PIs Schleider, J.; Crossfield, I.; Dressing, C.) alone has over 25,000 targets in \textit{K2} Campaigns 4 through 19, about five times larger than the number of M dwarfs observed with \Kepler. The ground based survey MEarth \citep{Nutzman2008} is surveying M dwarfs for planets, and the Search for habitable Planets EClipsing ULtra-cOOl Stars \citep[SPECULOOS;][]{Burdanov2017} is starting to search for planets around 1,000 late-type M dwarfs and brown dwarfs. Since small, short period planets are more prevalent around smaller stars, these surveys should be fruitful. The recently launched \textit{TESS} mission is also expected to find 500 to 1,000 planets around bright M dwarfs across the entire night sky \citep{Barclay2018,Ballard2018}. \citet{Barclay2018} predicted that 54 planets will be found with orbital periods less than 10 days around bright ($K_S < 10$) mid-type M dwarfs, assuming the occurrence rates of \citetalias{Dressing2015}. Based on our occurrence rates, we expect a more optimistic \textit{TESS} yield of $\sim$100 planets around bright mid-type M dwarfs.

\section{Acknowledgements}
\label{sec:acknowledgements}

We would like to thank WIYN observing assistants Amy Robertson, Anthony Paat, Christian Soto, Dave Summers, Doug Williams, Karen Butler, and Malanka Riabokin, DCT telescope operators Andrew Hayslip, Heidi Larson, Jason Sanborn, and Teznie Pugh, and IRTF telescope operators Brian Cabreira, Dave Griep, Eric Volquardsen, Greg Osterman, and Tony Matulonis. We would also like to thank NOAO observing support scientists Daryl Willmarth, Dianne Harmer, Mark Everett, and Susan Ridgway. K. Hardegree-Ullman would like to thank Adam Schneider and Wayne Oswald for conducting some of the observations herein at the DCT and WIYN, and Courtney Dressing, Gijs Mulders, and Jon Zink for informative discussions regarding planet occurrence rates and statistics.

Data presented herein were obtained at the WIYN Observatory from telescope time allocated to NN-EXPLORE through the scientific partnership of the National Aeronautics and Space Administration, the National Science Foundation, and the National Optical Astronomy Observatory. This work was supported by a NASA WIYN PI Data Award, administered by the NASA Exoplanet Science Institute.

These results made use of the Discovery Channel Telescope at Lowell Observatory. Lowell is a private, non-profit institution dedicated to astrophysical research and public appreciation of astronomy and operates the DCT in partnership with Boston University, the University of Maryland, the University of Toledo, Northern Arizona University and Yale University. The upgrade of the DeVeny optical spectrograph has been funded by a generous grant from John and Ginger Giovale.

Guoshoujing Telescope (the Large Sky Area Multi-Object Fiber Spectroscopic Telescope LAMOST) is a National Major Scientific Project built by the Chinese Academy of Sciences. Funding for the project has been provided by the National Development and Reform Commission. LAMOST is operated and managed by the National Astronomical Observatories, Chinese Academy of Sciences.

This research has made use of the NASA Exoplanet Archive, which is operated by the California Institute of Technology, under contract with the National Aeronautics and Space Administration under the Exoplanet Exploration Program.

This work has made use of data from the European Space Agency (ESA) mission {\it Gaia} (\url{https://www.cosmos.esa.int/gaia}), processed by the {\it Gaia} Data Processing and Analysis Consortium (DPAC, \url{https://www.cosmos.esa.int/web/gaia/dpac/consortium}). Funding for the DPAC has been provided by national institutions, in particular the institutions participating in the {\it Gaia} Multilateral Agreement.

This work made use of the \url{gaia-kepler.fun} crossmatch database created by Megan Bedell.

The Pan-STARRS1 Surveys (PS1) and the PS1 public science archive have been made possible through contributions by the Institute for Astronomy, the University of Hawaii, the Pan-STARRS Project Office, the Max-Planck Society and its participating institutes, the Max Planck Institute for Astronomy, Heidelberg and the Max Planck Institute for Extraterrestrial Physics, Garching, The Johns Hopkins University, Durham University, the University of Edinburgh, the Queen's University Belfast, the Harvard-Smithsonian Center for Astrophysics, the Las Cumbres Observatory Global Telescope Network Incorporated, the National Central University of Taiwan, the Space Telescope Science Institute, the National Aeronautics and Space Administration under Grant No. NNX08AR22G issued through the Planetary Science Division of the NASA Science Mission Directorate, the National Science Foundation Grant No. AST-1238877, the University of Maryland, Eotvos Lorand University (ELTE), the Los Alamos National Laboratory, and the Gordon and Betty Moore Foundation.

The authors wish to recognize and acknowledge the very significant cultural role and reverence that the summit of Maunakea has always had within the indigenous Hawaiian community. We are most fortunate to have the opportunity to conduct observations from this mountain.

\facilities{WIYN (Hydra), DCT (DeVeny), IRTF (SpeX), NASA Exoplanet Archive}

\software{dustmaps \citep{Green2018}, IDL Astronomy User's Library \citep{Landsman1993}, IRAF \citep{Tody1986,Tody1993}, numpy \citep{Oliphant2015}, scikit-learn \citep{Pedregosa2011}, Spextool \citep{Cushing2004}, xtellcor \citep{Vacca2003}}

\newpage
\bibliography{PlanetOccurrence}\setlength{\itemsep}{-2mm}

\newpage

\movetabledown=2in
\begin{rotatetable}
\begin{deluxetable*}{llcclrccrc}
\tablecaption{Kepler Mid-Type M Dwarf Planets\label{tab:planets}}
\tablewidth{0pt}
\tablehead{\colhead{Planet Name} & \colhead{Host Type} & \colhead{$R_{\star}$} & \colhead{$R_{p}$} & \colhead{$a/R_{\star}$} & \colhead{Period} & \colhead{$t_{\mathrm{dur}}$} & \colhead{Transit} & \colhead{$S_p$} & \colhead{References}\\
\colhead{} & \colhead{(This Work)} & \colhead{($R_{\odot}$)} & \colhead{($R_{\oplus}$)} & \colhead{} & \colhead{(days)} & \colhead{(hours)} & \colhead{Probability} & \colhead{($S_{\oplus}$)} & \colhead{}}
\startdata
Kepler-42 b & M3\,V & $0.174\pm0.005$ & $0.757\pm0.025$ & $14.46^{+0.43}_{-0.40}$ & $1.2137706\pm2.4$$\times$$10^{-7}$ & $0.57\pm0.04$ & $0.0719\pm0.0021$ & $26.86^{+3.31}_{-3.08}$ & 2\\
Kepler-42 c &  &  & $0.727\pm0.023$ & $7.49^{+0.23}_{-0.21}$ & $0.45328731\pm5.0$$\times$$10^{-8}$ & $0.44\pm0.03$ & $0.1385\pm0.0041$ & $99.78^{+12.16}_{-11.26}$ & 2\\
Kepler-42 d &  &  & $0.662\pm0.032$ & $19.26^{+0.59}_{-0.54}$ & $1.86511236\pm7.3$$\times$$10^{-7}$ & $0.56\pm0.09$ & $0.0537\pm0.0016$ & $15.13^{+1.88}_{-1.75}$ & 2\\
Kepler-1650 b & M3.5\,V & $0.364\pm0.010$ & $1.044\pm0.042$ & $10.88^{+0.34}_{-0.31}$ & $1.53818001\pm1.5$$\times$$10^{-6}$ & $1.05\pm0.11$ & $0.0942\pm0.0028$ & $42.67^{+5.59}_{-4.88}$ & 2\\
Kepler-1646 b & M4\,Ve & $0.374\pm0.011$ & $1.809\pm0.129$ & $21.86^{+0.71}_{-0.65}$ & $4.48558383\pm1.1$$\times$$10^{-5}$ & $1.53\pm0.07$ & $0.0478\pm0.0015$ & $9.57^{+1.29}_{-1.11}$ & 3\\
Kepler-446 b & M4\,V & $0.217\pm0.006$ & $1.357\pm0.073$ & $14.94^{+0.46}_{-0.44}$ & $1.565409\pm3.3$$\times$$10^{-6}$ & $0.67\pm0.09$ & $0.0708\pm0.0022$ & $20.62^{+2.62}_{-2.50}$ & 4\\
Kepler-446 c &  &  & $1.003\pm0.051$ & $23.23^{+0.71}_{-0.68}$ & $3.036179\pm5.5$$\times$$10^{-6}$ & $0.96\pm0.10$ & $0.0449\pm0.0014$ & $8.47^{+1.11}_{-1.00}$ & 4\\
Kepler-446 d &  &  & $1.227\pm0.063$ & $33.02^{+1.03}_{-0.95}$ & $5.148921\pm2.2$$\times$$10^{-5}$ & $0.93\pm0.13$ & $0.0318\pm0.0010$ & $4.18^{+0.56}_{-0.49}$ & 4\\
\enddata
\tablecomments{References: (1) \citet{Angelo2017}, (2) \citet{Mann2017}, (3) \citet{Morton2016}, (4) \citet{Muirhead2015}}
\end{deluxetable*}
\end{rotatetable}

\newpage

\startlongtable
\begin{deluxetable*}{lccclcccc}
\tablecaption{Stellar parameters for spectroscopically classified targets.\label{tab:spec}}
\tablewidth{0pt}
\tablehead{\colhead{KIC ID} & \colhead{Distance} & \colhead{$M_{K_S}$} & \colhead{Telescope$^a$} & \colhead{Spec. Type$^b$} & \colhead{$T_{\mathrm{eff}}$$^c$} & \colhead{[Fe/H]} & \colhead{$R_{\star}$} & \colhead{$M_{\star}$} \\ 
\colhead{} & \colhead{(pc)} & \colhead{(mag)} & \colhead{} & \colhead{} & \colhead{(K)} & \colhead{(dex)} & \colhead{($R_{\odot}$)} & \colhead{($M_{\odot}$)} } 

\startdata
01995305 & $139.882\pm1.253$ & $6.784\pm0.039$ & W/I & M3.5\,V & 3325 & $-0.426\pm0.100$ & $0.331\pm0.010$ & $0.320\pm0.010$ \\
02164791 & \nodata & \nodata & D/I & M4.5\,Ve & 3157 & $-0.166\pm0.099$ & $0.218\pm0.044$ & $0.182\pm0.064$ \\
02283749 & $291.795\pm6.182$ & $5.546\pm0.053$ & W & M4\,Ve & 3241 & \nodata & $0.516\pm0.017$ & $0.518\pm0.015$ \\
02693779 & $176.658\pm1.632$ & $6.308\pm0.038$ & W & M4\,V & 3241 & \nodata & $0.399\pm0.013$ & $0.394\pm0.011$ \\
02832720 & $106.036\pm5.989$ & $6.352\pm0.125$ & W/I & M4\,Ve & 3241 & $0.284\pm0.152$ & $0.398\pm0.020$ & $0.386\pm0.022$ \\
02834637 & $125.012\pm0.700$ & $6.406\pm0.023$ & W/I & M4\,Ve & 3241 & $0.063\pm0.073$ & $0.387\pm0.011$ & $0.378\pm0.009$ \\
03219046 & $63.177\pm0.308$ & $8.621\pm0.026$ & W & M5\,Ve & 3073 & \nodata & $0.166\pm0.005$ & $0.134\pm0.003$ \\
03232191 & \nodata & \nodata & W & M4\,V & 3241 & \nodata & $0.258\pm0.053$ & $0.226\pm0.075$ \\
\enddata
\tablenotetext{a}{D = DCT, I = IRTF, L = LAMOST, W = WIYN}
\tablenotetext{b}{Stars that exhibit H$\alpha$ emission from a by-eye analysis are classified with the peculiar flag `e' to denote emission.}
\tablenotetext{c}{We adopt a $\pm88\,\mathrm{K}$ uncertainty for all temperatures from our spectral type vs.\ $T_{\mathrm{eff}}$ fit (Section~\ref{sec:teff}).}
\tablenotetext{d}{Kepler-1582}
\tablenotetext{e}{Kepler-1649}
\tablenotetext{f}{Kepler-1650}
\tablenotetext{g}{Kepler-1646}
\tablenotetext{h}{Kepler-42}
\tablenotetext{i}{Kepler-446}
\tablenotetext{j}{Kepler-445}
\end{deluxetable*}

\startlongtable
\begin{deluxetable*}{ccccccc}
\tablecaption{Stellar parameters for photometrically classified targets.\label{tab:phot}}

\tablehead{\colhead{KIC ID} & \colhead{Distance} & \colhead{$M_{K_S}$} & \colhead{Spec. Type} & \colhead{$T_{\mathrm{eff}}$$^a$} & \colhead{$R_{\star}$} & \colhead{$M_{\star}$} \\ 
\colhead{} & \colhead{(pc)} & \colhead{(mag)} & \colhead{} & \colhead{(K)} & \colhead{($R_{\odot}$)} & \colhead{($M_{\odot}$)} } 

\startdata
01433760 & $192.959\pm2.769$ & $6.697\pm0.046$ & M3\,V & 3410 & $0.348\pm0.012$ & $0.333\pm0.010$ \\
02831828 & $106.525\pm1.537$ & $8.923\pm0.056$ & M5\,V & 3073 & $0.148\pm0.005$ & $0.119\pm0.004$ \\
02971472 & $40.706\pm0.059$ & $6.979\pm0.019$ & M5\,V & 3073 & $0.313\pm0.009$ & $0.292\pm0.007$ \\
02983661 & $148.672\pm1.211$ & $6.832\pm0.029$ & M3\,V & 3410 & $0.331\pm0.010$ & $0.313\pm0.008$ \\
03101838 & $201.853\pm2.262$ & $5.846\pm0.032$ & M3\,V & 3410 & $0.468\pm0.014$ & $0.469\pm0.012$ \\
03228276 & $111.970\pm0.562$ & $6.865\pm0.025$ & M3\,V & 3410 & $0.327\pm0.010$ & $0.308\pm0.008$ \\
03443923 & $144.103\pm0.963$ & $5.575\pm0.026$ & M4\,V & 3241 & $0.511\pm0.015$ & $0.513\pm0.013$ \\
03545104 & $45.524\pm0.073$ & $7.464\pm0.019$ & M5\,V & 3073 & $0.260\pm0.008$ & $0.231\pm0.006$ \\
\enddata
\tablenotetext{a}{We adopt a $\pm88\,\mathrm{K}$ uncertainty for all temperatures from our spectral type vs.\ $T_{\mathrm{eff}}$ fit (Section~\ref{sec:teff}).}
\end{deluxetable*}

\end{document}